\documentclass[10pt,aps,prd,twocolumn,showpacs,superscriptaddress,eqsecnum,longbibliography,nofootinbib]{revtex4-2}
\usepackage{mathrsfs,amsmath,amsthm,latexsym,amssymb,amsfonts,epsfig,cancel,enumerate,graphicx,subfigure,txfonts}
\usepackage{xcolor}
\definecolor{navy}{RGB}{0,0,150}
\usepackage[colorlinks, linkcolor=blue, citecolor=blue, urlcolor=blue, plainpages=false, pdfstartview=FitH]{hyperref}
\usepackage{appendix}
\allowdisplaybreaks
\newcommand{\GZU}{School of Physics, Guizhou University, Guiyang 550025, China}
\newcommand{\UOW}{Faculty of Physics, University of Warsaw, Pasteura 5, 02-093 Warsaw, Poland}
\newcommand{\UOE}{Department Physik, Institut f\"ur Quantengravitation, Theoretische Physik III, Friedrich-Alexander-Universit\"at Erlangen-Nürnberg, Staudtstra{\ss}e 7/B2, 91058 Erlangen, Germany}
\newcommand{\BNU}{Department of Physics, Beijing Normal University, Beijing 100875, China}

\begin{document}

\title{Shadow and stability of quantum-corrected black holes}

%-------------------------------------------------------------
\author{Jinsong Yang}
\email{jsyang@gzu.edu.cn}
\affiliation{\GZU}

\author{Cong Zhang}
\email{cong.zhang@gravity.fau.de}
\affiliation{\UOE}
\affiliation{\UOW}

\author{Yongge Ma}
\thanks{Corresponding author}
\email{mayg@bnu.edu.cn}
\affiliation{\BNU}
%-------------------------------------------------------------

%-------------------------------------------------------------

\begin{abstract}

Recently the quantum Oppenheimer-Snyder gravitational collapse model has been proposed in loop quantum gravity, providing quantum-corrected Schwarzschild spacetimes as the exterior of the collapsing dust ball. In this paper, the quantum gravity effects on the black hole shadows in this model are studied, and the stability of the quantum-corrected black holes is also analyzed by calculating the quasinormal modes. It turns out that the quantum correction always shrinks the radius of shadows, and the quantum-corrected black holes are stable against the scalar and vector perturbations.

\end{abstract}

%-------------------------------------------------------------
\maketitle
%-------------------------------------------------------------
%\tableofcontents
%-------------------------------------------------------------

\section{Introduction}

The singularity theorems in general relativity (GR) predicts the formation of black holes (BHs) and spacetime singularities~\cite{Penrose:1964wq,Hawking:1970zqf}. The appearance of singularities implies the breakdown of GR when spacetime curvature increases unboundedly. Therefore, it is nature to expect that some quantum theory of gravity becomes domain in these regions and can cure the singularities. Up to now, some competitive quantum theories of gravity have been proposed. Among them, loop quantum gravity (LQG), a background-independent and nonperturbative approach to quantum gravity, has received considerable attention and has been widely investigated (see, e.g.,~\cite{Rovelli:2004tv,Thiemann:2007pyv} for books, and~\cite{Thiemann:2002nj,Ashtekar:2004eh,Han:2005km,Giesel:2012ws,Rovelli:2011eq,Perez:2012wv} for articles). After thirty years of theoretical research, the canonical (Hamiltonian) and the covariant (Lagrangian) formulations of LQG have achieved individual and remarkable successes. Besides the predictions of discretized geometries and the interpretation of BH entropy in the canonical formulation~\cite{Rovelli:1994ge,Ashtekar:1996eg,Ashtekar:1997fb,Yang:2016kia,Thiemann:1996at,Ma:2010fy,Ashtekar:1997yu,Song:2020arr}, well-defined dynamical descriptions in both formulations were proposed, and the consistency between them was checked in certain senses~\cite{Thiemann:1996aw,Thiemann:1997rt,Yang:2015zda,Alesci:2015wla,Zhang:2018wbc,Zhang:2019dgi,Engle:2007wy,Freidel:2007py,Alesci:2011ia,Yang:2021den}. Gravity coupled to matters were studied and revisited in LQG, leading to the resolution of some long-standing problems~\cite{Thiemann:1997rt,Lewandowski:2021bkt,Zhang:2022bzp}. Furthermore, loop quantization program has been successfully extended to high-dimensional theories of gravity as well as alternative gravitational theories~\cite{Bodendorfer:2011nx,Han:2013noa,Long:2019nkf,Long:2020wuj,Long:2020agv,Zhang:2011vi,Zhang:2011qq,Zhang:2011vg,Zhang:2011gn,Ma:2011aa,Zhou:2012ie,Chen:2018dqz,Zhang:2020smo}. Aiming to test the ideas and techniques of full LQG and to quantize the symmetry-reduced phase space of the theory, loop quantum cosmology (LQC) and loop quantum black hole (LQBH) have been studied, and substantial progresses have been made in solving the cosmological big-bang singularity and the BH singularity~\cite{Ashtekar:2003hd,Ashtekar:2006rx,Ashtekar:2006wn,Ding:2008tq,Yang:2009fp,Assanioussi:2018hee,Li:2018opr,Ashtekar:2005qt,Gambini:2020nsf}. Although a systematical derivation of these symmetric-reduced models from LQG is still absent up to now, some progress has been made on the relation between LQG and these models by calculating the expectation value of the Hamiltonian in LQG under certain coherent state peaked at some point in the classical phase space~\cite{Han:2019vpw,Han:2019feb,Liegener:2020dcg,Zhang:2020mld,Zhang:2021qul}.

The spherically symmetric gravitational collapse plays an important role in understanding the formation of BH and its singularity in GR. Classically, the gravitational collapse model was constructed by Oppenheimer and Snyder~\cite{Oppenheimer:1939ue} (OS model). In this model, the interior sourced matter is assumed to be a spherically symmetric and homogeneous pressureless dust, and thus it can be modeled by a FRW model. Due to the simplicity, this model is exactly solvable, providing us with a new window to understand more complicated and realistic dynamical processes of gravitational collapse. There is no doubt that to have a complete description of the gravitational collapse, one needs to incorporate the quantum gravity effects. In this direction, the quantum effects of LQG on the gravitational collapse models are being studied~\cite{Bojowald:2005qw,Bojowald:2009ih,Marto:2013soa,Kelly:2020lec,BenAchour:2020bdt,BenAchour:2020gon,Munch:2020czs,Munch:2021oqn,Husain:2021ojz,Giesel:2021dug}. In particular, an effective Hamiltonian and an effective metric for the vaccuum exterior solution for the OS collapse model were derived in the effective context of spherically symmetric spacetime~\cite{Kelly:2020lec}. An interior to exterior approach for quantum OS model has been proposed recently in~\cite{Lewandowski:2022zce}, where the effective interior spacetime with LQC corrections is carried out to its exterior one by certain matching condition on its boundary surface. It turns out that in preliminary stage the effective exterior spacetime is a quantum-corrected Schwarzschild spacetime. As collapse goes on, a quantum-corrected BH forms with the occurrence of horizon. In the late stage, the collapsing phase will be bounced to an expanding phase, resulting in a transformation from a BH to a white hole. Moreover the global causal structure of the maximal extension of the quantum-corrected Schwarzschild spacetime has been studied in~\cite{Lewandowski:2022zce}.

The successful detection of gravitational-wave from a pair of BHs and the observation of the images of supermassive BH are promoting strongly the observational investigation of BHs~\cite{LIGOScientific:2017bnn,EventHorizonTelescope:2019dse,EventHorizonTelescope:2022wkp}. On one hand, the systematical analysis of BH shadow and photon ring has been proposed \cite{Bardeen:1973,Gralla:2019xty}, providing a possibility to distinguish various BHs in different gravitational theories. On the other hand, it is widely believed that the study of quasinormal modes (QNMs) of BH plays important roles not only in analyzing the stability of BHs, but also in understanding the gravitational wave signals \cite{Regge:1957td,Chandrasekhar:1975zza,Gundlach:1993tp}. These works attract much more attentions in the fields of both astrophysics and theoretical physics. In this paper, we will study how the quantum correction to a BH spacetime affects the shadow and QNMs of the BH, by comparing them to those of the classical Schwarzschild BH. We are going to reveal the effects from the effective exterior spacetime, though there are also other quantum gravity effects in BH image \cite{Zhang:2023okw}.

The rest of this paper is organized as follows. In Sec.~\ref{sec-II}, we recall an interior to exterior approach for the OS model to obtain the effective exterior spacetime. In Sec.~\ref{sec-III}, we study the quantum gravitational effects on shadows and QNMs of BHs. Our results are summarized in Sec.~\ref{sec-IV}. The detailed derivation of the junction conditions in the main text will be presented in Appendix~\ref{appendix}.

\section{The strategy to generate exterior solutions}\label{sec-II}

In this section, we recall the OS model. In this model, the spacetime manifold ${\cal M}$ is divided into two regions, the interior region ${\cal M}^-$ and the exterior region ${\cal M}^+$, by its timelike boundary 3-surface $\Sigma={\cal M}^-\cap{\cal M}^+$. The former is assumed to be a FRW spacetime region sourced with a spherical dust ball (cloud) with homogenous density with the following line element
\begin{align}\label{eq:line-element-interior}
  {\rm d}s_-^2=-{\rm d}\tau^2+a(\tau)^2\left[\frac{{\rm d}\tilde{r}^2}{1-k\tilde{r}^2}+\tilde{r}^2\,{\rm d}\Omega^2\right],
\end{align}
where $\tau$ denotes the proper time of the comoving observer, $a(\tau)$ is the scale factor, ${\rm d}\Omega^2\equiv{\rm d}\theta^2+\sin^2\theta\,{\rm d}\phi^2$, and $k=-1,0,+1$ for the open, flat and closed FRW models, respectively. The boundary surface $\Sigma$ of ${\cal M}^-$ is located at a constant $\tilde{r}=\tilde{r}_0$, in the comoving coordinates, and can be parametrized by $(\tau,\theta,\phi)$.  While the latter is assumed to be a static and spherically symmetric spacetime region with the line element, and can be expressed in Schwarzschild coordinates $(t,r,\theta,\phi)$ as
\begin{align}\label{eq:line-element-exterior}
  {\rm d}s_+^2=-f(r){\rm d}t^2+f(r)^{-1}{\rm d}r^2+r^2{\rm d}\Omega^2.
\end{align}
To match these two regions at $\Sigma$, suitable boundary conditions need to be imposed. In the classical theory, the conditions have been studied by Darmois and Israel~\cite{Darmois:1927,Israel:1966rt}, and hence are called as the Darmois-Israel junction conditions. By matching the exterior to the interior of dust collapsing star via the Darmois-Israel junction conditions, the exterior solutions could be generated dynamically by the interior solutions. It turns out that, by matching ${\cal M}^+$ to ${\cal M}^-$ along $\Sigma$ generated by geodesics, the junction conditions lead to the following relations between the two regions~\cite{Poisson:2004bk,Lewandowski:2022zce} (see Appendix~\ref{appendix} for details)
\begin{align}
a(\tau)\tilde{r}_0&=r(\tau)\label{eq:match-1},\\
f(r)&=\left(1-k\tilde{r}_0^2\right)-H(\tau)^2r^2,\label{eq:match-2}
\end{align}
where $H(\tau)\equiv\frac{1}{a(\tau)}\frac{{\rm d}a(\tau)}{{\rm d}\tau}$ is the Hubble parameter.

Now we impose the dynamical equation for the interior region ${\cal M}^-$. In the classical theory, the dynamics of the interior spacetime satisfies the Friedmann equation
\begin{align}\label{eq:classical-Friedmann-eq}
H^2=\frac{8\pi G}{3}\rho-\frac{k}{a^2},
\end{align}
where $G$ is the Newtonian gravitational constant, $\rho(\tau)=\frac{M}{\frac{4\pi}{3}r(\tau)^3}$ is the energy density in ${\cal M}^-$. Inserting Eq.~\eqref{eq:classical-Friedmann-eq} into Eq.~\eqref{eq:match-2} and using Eq.~\eqref{eq:match-1}, we have
\begin{align}\label{eq:G-r-classical}
f(r)&=\left(1-k\tilde{r}_0^2\right)+k\frac{r^2}{a^2}-\frac{8\pi G}{3}\rho r^2\notag\\
&=1-\frac{8\pi G}{3}\rho r^2\notag\\
&=1-\frac{R_s}{r},
\end{align}
where $R_s\equiv 2GM$ denotes Schwarzschild radius. Hence the exterior metric is nothing but Schwarzschild metric as expected, i.e.,
\begin{align}\label{eq:line-element-exterior-classical}
  {\rm d}s_+^2&=-\left(1-\frac{R_s}{r}\right){\rm d}t^2+\left(1-\frac{R_s}{r}\right)^{-1}{\rm d}r^2+r^2{\rm d}\Omega^2.
\end{align}

We assume that the Darmois-Israel junction conditions be still valid for the effective theory, which is widely adopted in the literature~\cite{Bojowald:2005qw,Munch:2020czs,Piechocki:2020bfo}. In addition, to achieve the aim that the quantum effects in the interior can be carried into the exterior, one needs also to specify the interior effective dynamical equation. The effective Friedmann equation reads~\cite{Ashtekar:2006wn,Vandersloot:2006ws,Vandersloot:2006gga}
\begin{align}\label{eq:Friedmann-eff}
  H_{\rm eff}^2=\frac{8\pi G}{3}\left(\rho-\frac{3}{8\pi G}\frac{k}{a^2}\right)\left(1-\frac{\rho-\frac{3}{8\pi G}\frac{k}{a^2}}{\rho_c}\right),
\end{align}
where $\rho_c=\frac{3}{8\pi G\gamma^2\Delta}$ is the critical energy density. The parameter $\Delta=4\sqrt{3}\pi\gamma\ell^2_{\rm p}$ is the area gap, where $\ell^2_{\rm p}=G\hbar$ and $\gamma$ denotes the Immirzi parameter whose value has been fixed as $0.2375$ by calculating BH entropy~\cite{Domagala:2004jt,Meissner:2004ju}. Note that the resulting effective Friedmann equation \eqref{eq:Friedmann-eff} is suitable for arbitrary matter sources though it is derived with a massless scalar field. The reason behind this is that the quantum corrections to Friedmann equation arise completely from the quantum modification in the gravitational part. We now consider the dynamics of the interior spacetime determined by the effective equation \eqref{eq:Friedmann-eff}. Therefore, repeating the matching procedure, the quantum-corrected exterior metric can be obtained as \cite{Lewandowski:2022zce}
\begin{align}\label{eq:exterior-metric}
{\rm d}s_{+,{\rm eff}}^2&=-f(r){\rm d}t^2+f(r)^{-1}{\rm d}r^2+r^2{\rm d}\Omega^2,
\end{align}
where
\begin{align}\label{eq:f-effective}
 f(r)&=1-\frac{R_s}{r}+\frac{\gamma^2\Delta}{r^2}\left(\frac{R_s}{r}-k\tilde{r}_0^2\right)^2.
\end{align}
It is easy to see that the effective metric \eqref{eq:exterior-metric} tends to the classical metric \eqref{eq:line-element-exterior-classical} as $\Delta\rightarrow0$. This behavior also holds for large scale with $r>>\sqrt{\Delta}$. Hence the effective exterior metric \eqref{eq:exterior-metric} goes to the classical Schwarzschild metric \eqref{eq:line-element-exterior-classical}. It is remarkable that the effective exterior metric with $k=0$ in Eq.~\eqref{eq:exterior-metric} is just the one obtained directly by loop quantizing the spherically symmetric exterior spacetime in~\cite{Kelly:2020uwj}. In contrast to the classical case where the exterior metric always takes the Schwarzschild one for different $k$, the quantum-corrected exterior metrics for different $k$ are actually different from each other in the effective theory.

The energy density $\rho_+$ in ${\cal M}^+$ measured by the static observer can be obtained as
\begin{align}\label{eq:energydensity-Hubble}
 \rho_+(r)=-\frac{1}{8\pi G}\frac{G_{tt}}{g_{tt}}=\frac{1}{8\pi G}\left\{k\frac{\tilde{r}_0^2}{r^2}+H(r)\left[3H(r)+2rH'(r)\right]\right\},
\end{align}
where $G_{tt}$ and $g_{tt}$ denote the $tt$-components corresponding to the Einstein tensor $G_{ab}$ and the metric tensor $g_{ab}$, respectively. Here the Hubble parameter $H$ should be understood as a function of $\rho=\frac{M}{\frac{4\pi}{3}r^3}$, and thus as a function of $r$, by using the Friedmann equation. In the classical case, the vanishing energy density determined by Eq.~\eqref{eq:energydensity-Hubble} indicates the the vaccum exterior. A straightforward calculation shows that the effective energy density reads
\begin{align}
 \rho_{+}^{\rm eff}=\frac{3}{8\pi G}\frac{\gamma^2\Delta}{r^4}\left(\frac{R_s}{r}-k\tilde{r}_0^2\right)\left(\frac{R_s}{r}-\frac{1}{3}k\tilde{r}_0^2\right).
\end{align}

\begin{figure}
  \includegraphics[width=\columnwidth]{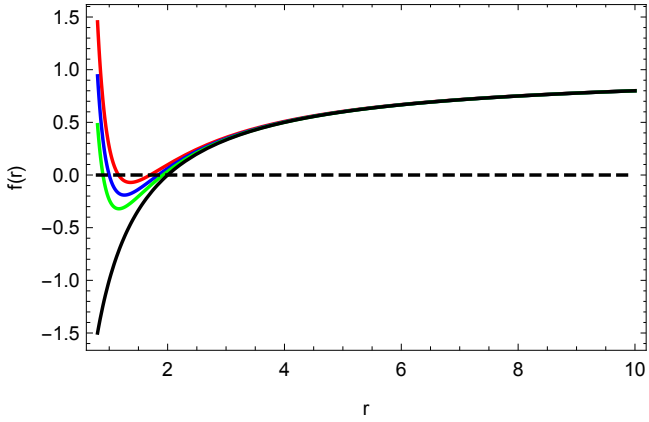}
  \caption{Plots of $f(r)$ with respect to $r$ for Schwarzschild case (black) and the effective cases corresponding to $k=-1$ (red), $k=0$ (blue) and $k=+1$ (green), with $\gamma=1,R_s=2$, $\Delta=0.25$, and $\tilde{r}_0=0.5$.}
  \label{fig:f-r}
\end{figure}

As collapse goes on, a quantum-corrected BH will form with the occurrence of horizon for a large $M$. Note that the exterior metric \eqref{eq:exterior-metric} is static. Hence the vector field $\xi^a=\left(\frac{\partial}{\partial t}\right)^a$ is a Killing vector field. By definition, the Killing horizon ${\cal K}$ is a null hypersurface determined by $\left.\xi^a\xi_a\right|_{\cal K}=0$. Therefore, the Killing horizon corresponds to the surface with radius $r$ satisfying $f(r)=0$. The functions $f(r)$ of the quantum-corrected BHs are compared to that of Schwarzschild BH in Fig.~\ref{fig:f-r}. Interestingly, it is easy to see from Fig.~\ref{fig:f-r} that each of the quantum-corrected BHs has two Killing horizons at $f(r)=0$, while there is only one horizon in the BH of classical theory. It turns out that the three horizons, the event, Killing and apparent horizons, are the same for the spacetime with metric \eqref{eq:exterior-metric}.
%================fig================
%===================================

\begin{figure}
  \includegraphics[width=\columnwidth]{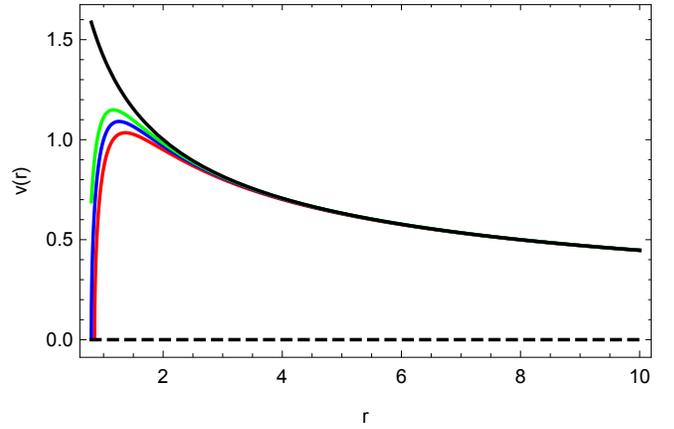}
  \caption{Plots of $v(r)$ with respect to $r$ for Schwarzschild case (black) and the effective cases corresponding to $k=-1$ (red), $k=0$ (blue) and $k=+1$ (green), with $\gamma=1,R_s=2$, $\Delta=0.25$, and $\tilde{r}_0=0.5$.}
  \label{fig:v-r}
\end{figure}

In the late stage of the evolution in the effective LQC model, the collapse of the dust cloud will stop whence $v(r)\equiv\dot{r}=0$ is satisfied, which leads to $f(r)=1-k\tilde{r}_0^2$. At this point, $r$ takes the minimum value
\begin{align}
r_{\rm m}=\left(x\gamma^2\Delta\right)^{1/3}-k\tilde{r}_0^2\frac{\left(x\gamma^2\Delta\right)^{2/3}}{3x},
\end{align}
where $x\equiv\frac12\left(R_s+\sqrt{R_s^2+\frac{4}{27}k^3\tilde{r}_0^6\gamma^2\Delta}\right)$. Thus the exterior effective metric is well defined for $r\geqslant r_{\rm m}$. In the limit $r\rightarrow r_{\rm m}$, the interior (equivalent) energy density $\rho-\frac{3}{8\pi G}\frac{k}{a^2}$ of the dust cloud reaches its maximum value $\rho_c$. The functions $v(r)$ are plotted for the Schwarzschild and quantum-corrected cases in Fig.~\ref{fig:v-r}. It is shown that, during the collapse with decreasing $r$, the collapsing velocity of the boundary surface increases to its maximum and then decreases to 0 in the effective theory, while it increases to infinity in the classical case.
%================fig================
%===================================
After that, the collapsing phase will be bounced to an expanding phase in the effective theory, resulting in a transformation from a BH to something similar to a white hole. Therefore, the quantum gravitational effects resolve the classical BH singularity and replace it by a quantum bounce at small scale, and the effective metric agrees quite well with the Schwarzschild metric at large scale. The global causal structure of the maximally extended BH spacetime with $k=0$ in the effective theory has been studied in ~\cite{Lewandowski:2022zce}.

\section{The quantum gravitational effects}\label{sec-III}

In this section, we will study the quantum gravity effects on observables such as the BH shadow, the QNMs and the ringdown waveform, and compare them to those of Schwarzschild case.

\subsection{Shadows, rings and lensing rings}\label{sec-III-A}

\begin{figure*}[tb]
\includegraphics[scale=0.75]{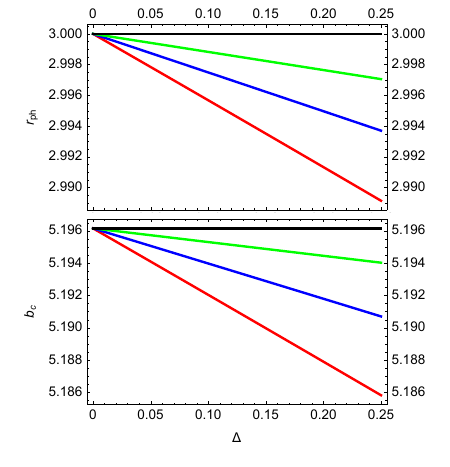}\quad
\includegraphics[scale=0.75]{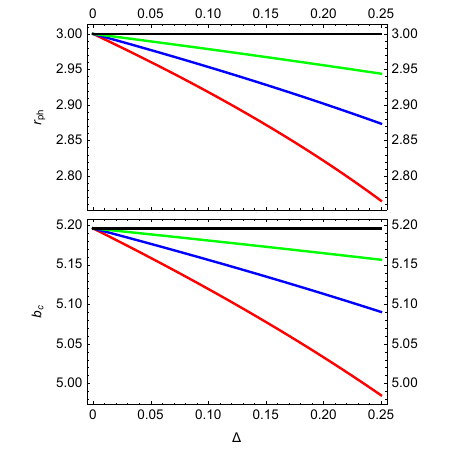}\quad
\includegraphics[scale=0.75]{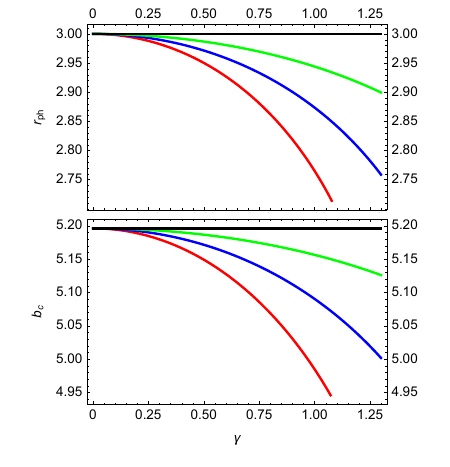}
\caption{Behavior of $r_{\rm ph}$ and $b_c$ for Schwarzschild BH (black) and quantum-corrected BHs corresponding to $k=-1$ (red), $k=0$ (blue) and $k=+1$ (green): In the left and middle panels, we fix $\gamma=0.2375$ and $\gamma=1$, respectively, by varying $\Delta$. In the right panel, we fix $\Delta=0.25$ and vary $\gamma$. The parameters $R_s$ and $\tilde{r}_0$ are taken to $R_s=2$ and $\tilde{r}_0=0.5$.}\label{fig:photon-sphere}
\end{figure*}

Considering the spherically symmetric BHs, the trajectory of a light outside a BH always lies on a plane. Taking into account the conserved energy $E$ and angular momentum $J$, the orbital equation of light ray approaching to the BHs described by Eq.~\eqref{eq:exterior-metric} can be obtained as an ordinary differential equation of the radius $r$ in terms of the azimuthal angle $\phi$ on the orbital plane as
\begin{align}\label{eq:orbital-eq-1}
 \left(\frac{{\rm d}r}{{\rm d}\phi}\right)^2=r^4\left(\frac{1}{b^2}-\frac{f(r)}{r^2}\right)\equiv \bar{V}(r),
\end{align}
where $b=J/E$ is the impact parameter associated to the light ray. Hence the trajectory of light ray is completely determined by its impact parameter $b$. The radius $r_{\rm ph}$ of the photon sphere formed by a bounded orbit of light is determined by
\begin{align}
 \left.\bar{V}(r)\right|_{r=r_{\rm ph}}&=0,\label{eq:photon-sphere-1}\\
 \left.\frac{{\rm d}\bar{V}(r)}{{\rm d}r}\right|_{r=r_{\rm ph}}&=0.\label{eq:photon-sphere-2}
\end{align}
Due to the spherical symmetry, Eq.~\eqref{eq:photon-sphere-2} can be reduced to
\begin{align}
 \left.\frac{{\rm d}}{{\rm d}r}\frac{f(r)}{r^2}\right|_{r=r_{\rm ph}}=0.\label{eq:photon-sphere-2-2}
\end{align}
Inserting the solution $r=r_{\rm ph}$ of Eq.~\eqref{eq:photon-sphere-2-2} into Eq.~\eqref{eq:photon-sphere-1}, one obtains the critical compact parameter $b_c$ corresponding to the photon sphere as
\begin{align}
 b_c=\frac{r_{\rm ph}}{\sqrt{f(r_{\rm ph})}}.
\end{align}
According to the value of $b$, the light trajectory near a BH can be classified into the following three situations: (i) $b=b_c$, the light ray will surround BHs in the circular orbit; (ii) $b<b_c$, the light ray will approach and fall into the BH; (iii) $b>b_c$, the light ray will be binded by the BH and then escapes to spatial infinity. For the Schwarzschild BH, the orbital radius of photon sphere and the corresponding impact parameter read respectively as
\begin{align}
 r^{\rm Sch}_{\rm ph}=3\frac{R_s}{2}, \qquad b^{\rm Sch}_c=3\sqrt{3}\frac{R_s}{2}.
\end{align}
For the quantum-corrected BHs, these physical quantities depend on $\Delta$ as well as $\gamma$. It should be noted that, in Fig.~\ref{fig:f-r} and the following numerical calculations, the parameter $R_s$ is fixed as $R_s=2$, and hence the impact parameter $b$ is scaled by $GM$. The parameters $r_{\rm ph}$ and $b_c$ for different values of parameters $\gamma$ and $\Delta$ are plotted in Fig.~\ref{fig:photon-sphere} for the quantum-corrected BHs and compared to those of the Schwarzschild BH. In the left and middle panels of Fig.~\ref{fig:photon-sphere}, the Immirzi parameter is chosen as $\gamma=0.2375$ and $\gamma=1$, respectively. It is clear that as $\Delta$ increases, both $r_{\rm ph}$ and $b_c$ monotonically decrease for each of the quantum-corrected BHs. The right panel of Fig.~\ref{fig:photon-sphere} shows that both $r_{\rm ph}$ and $b_c$ have a similar behavior with fixed $\Delta=0.25$ and varying $\gamma$. It indicates that for small $\Delta$, both $r_{\rm ph}$ and $b_c$ for each of the quantum-corrected BHs are always smaller than those of Schwarzschild BH with the same mass. It should be noted that by choosing $\Delta=0.25$, its effect on the shadows of BHs are obviously different for different theories, although $\Delta$ should be a very small number. Also, we will fix $\gamma=1$ for simplicity in the following.
%================fig================
%===================================

To study the light bending near a BH, it is convenient to introduce a variable $u=1/r$. Then the orbital equation \eqref{eq:orbital-eq-1} can be expressed in terms of $u$ as
\begin{align}\label{eq:orbit-eq-u-phi}
 \left(\frac{{\rm d}u}{{\rm d}\phi}\right)^2=\frac{1}{b^2}-f\left(\frac1u\right)u^2.
\end{align}
Hence the total change in azimuthal angle outside the horizon of the trajectory for the situation (ii) with $b<b_c$ is given by~\cite{Gralla:2019xty,Peng:2020wun}
\begin{align}\label{eq:totalangle1}
 \phi=\int^{u_h}_0\frac{1}{\sqrt{\frac{1}{b^2}-f\left(\frac1u\right)u^2}}{\rm d}u,
\end{align}
where $u_h:=1/r_h$ with $r_h$ being the radius of the (outermost) horizon. In the situation (iii) with $b>b_c$, the total change in angle reads
\begin{align}\label{eq:totalangle2}
 \phi&=2\int^{u_{\rm max}}_0\frac{1}{\sqrt{\frac{1}{b^2}-f\left(\frac1u\right)u^2}}{\rm d}u,
\end{align}
where $u_{\rm max}:=1/r_{\rm min}$ with $r_{\rm min}$ being the light ray's radial minimal distance from its trajectory to the BH. Let $n=\phi/(2\pi)$ be the total number of orbits, which is a function of $b$, satisfying~\cite{Gralla:2019xty,Peng:2020wun}
\begin{align}\label{eq:nb-m}
 n(b)=\frac{2m-1}{4},\qquad m=1,2,3,\cdots.
\end{align}
For each given $m$, there will be two solutions for Eq.~\eqref{eq:nb-m}~\cite{Gralla:2019xty,Peng:2020wun}, denoted by $b_m^\pm$ with $b_m^-$ and $b_m^+$ being the minimum and the maximum solutions, respectively. Then the rays can be classified as follows:
\begin{enumerate}[(1)]
 \item Direct: $n<3/4$ $\quad\Leftrightarrow\quad$ $b\in(0,b_2^-)\cup(b_2^+,\infty)$;
 \item Lensed: $3/4<n<5/4$ $\quad\Leftrightarrow\quad$ $b\in(b_2^-,b_3^-)\cup(b_3^+,b_2^+)$;
 \item Photon ring: $n>5/4$ $\quad\Leftrightarrow\quad$ $b\in(b_3^-,b_3^+)$.
\end{enumerate}
Note that the orbit equation of the time-like geodesic reads,
\begin{align}
 \left({\frac{{\rm d}r}{{\rm d}\phi}}\right)^2=r^4\left(\frac{1}{b^2}-\frac{f(r)}{r^2}-\frac{f(r)}{J^2}\right)\equiv \tilde{V}(r).
\end{align}
The radius $r_{\rm isco}$ of the innermost stable circular orbit (isco) is determined by
\begin{align}
 \left.\tilde{V}(r)\right|_{r=r_{\rm isco}}=0, \quad\left.\frac{{\rm d}\tilde{V}(r)}{{\rm d}r}\right|_{r=r_{\rm isco}}=0,\quad\left.\frac{{\rm d}^2\tilde{V}(r)}{{\rm d}r^2}\right|_{r=r_{\rm isco}}=0.
\end{align}

\begin{figure*}
\includegraphics[scale=1]{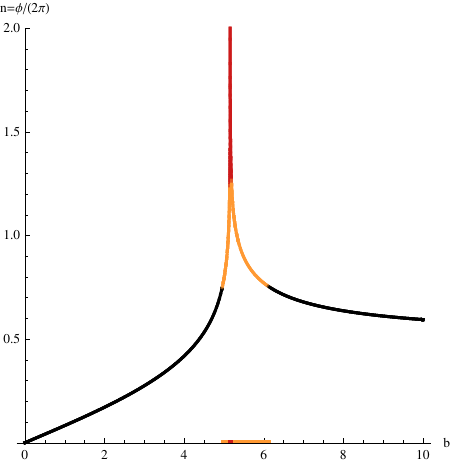}\qquad\qquad
\includegraphics[scale=0.6]{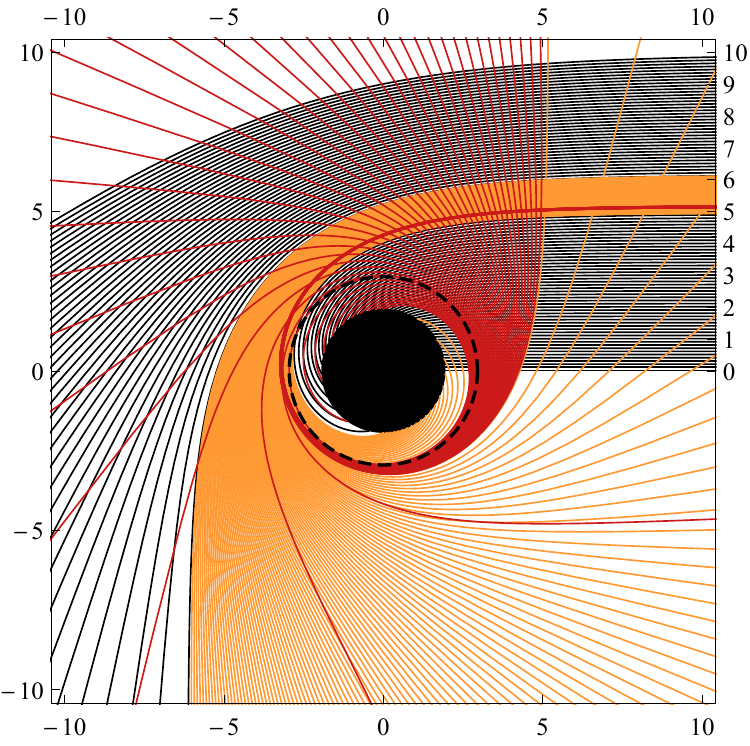}
\caption{Behavior of photons in the effective spacetime with $k=0$ as a function of $b$: On the left panel, the total number of orbits, $n=\phi/(2\pi)$, is shown. The black, gold and red lines correspond to the direct, lensed and photon rings, respectively.  On the right panel, a selection of associated trajectories in the Euclidean polar coordinates $(r,\phi)$ is plotted. The impact parameter spacing is $0.1$, $0.01$ and $0.001$ in the direct (black), lensed (gold) and photon rings (red), respectively. The BH is represented by a black disk, while the circular orbit of light is shown as a dashed black circular. The parameters are $R_s=2$, $\gamma=1$, and $\Delta=0.25$.}\label{fig:trajectory}
\end{figure*}

%================table================
\begin{table*}
\caption{Various involved physical quantities for Schwarzschild spacetime (Sch) and the quantum-corrected spacetimes corresponding to $k=-1$ (QC-I), $k=0$ (QC-II) and $k=+1$ (QC-III), with $R_s=2$, $\gamma=1$, $\Delta=0.25$, and $\tilde{r}_0=0.5$.}
\begin{ruledtabular}
\begin{tabular}{cccccccccc}
  BHs & $r_h$   & $r_{\rm ph}$   &$r_{\rm isco}$& $b_c$  & $b_1^-$ & $b_2^-$ & $b_2^+$ & $b_3^-$ & $b_3^+$ \\
  \hline
  Sch    & 2            & 3            & 6            & 5.19615 & 2.84770 & 5.01514 & 6.16757 & 5.18781 & 5.22794 \\
  QC-I   & 1.70143 & 2.76504 & 5.65822 & 4.98500 & 2.47085 & 4.69676 & 6.06651 & 4.96522 & 5.03312\\
  QC-II  & 1.83929 & 2.87357 & 5.83769 & 5.09041 & 2.64698 & 4.85475 & 6.12553 & 5.07672 & 5.13117\\
  QC-III & 1.91810 & 2.94392 & 5.95443 & 5.15643 & 2.74724 & 4.94618 & 6.15888 & 5.14542 & 5.19279 \label{table:1}
\end{tabular}
\end{ruledtabular}
\end{table*}
%===================================
In Table \ref{table:1}, the involved physical quantities are shown for the classical and quantum-corrected BHs. It indicates that for the fixed parameters $R_s=2$, $\gamma=1$, $\Delta=0.25$, and $\tilde{r}_0=0.5$, the quantum effect always shrinks the corresponding quantities. It is worth noting that the universal conjecture~\cite{Lu:2019zxb,Feng:2019zzn}
\begin{align}
 \frac32r_h\leqslant r_{\rm ph} \leqslant \frac{b_c}{\sqrt{3}} \leqslant 3GM
\end{align}
is satisfied for Schwarzschild BHs as well as quantum-corrected BHs. The behaviors of photons in the effective spacetime of the quantum-corrected BH with $k=0$ are plotted in Fig.~\ref{fig:trajectory}. The left panel of Fig.~\ref{fig:trajectory} depicts the total number of orbits, and the right panel shows the trajectories of light rays surrounding the quantum-corrected BH.
%================fig================
%===================================

\begin{figure*}
\includegraphics[scale=1]{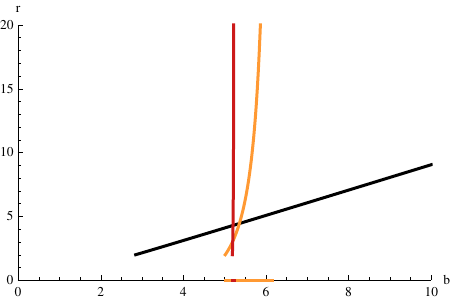}\hspace{2cm}
\includegraphics[scale=1]{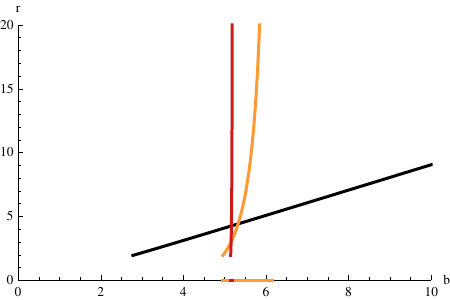}
\caption{The first three transfer functions for a face-on thin disk in Schwarzschild spacetime (left) and the effective spacetime with $k=0$ (right), representing the radial coordinate of the first (black), second (gold), and third (red) intersections with a face-on thin disk outside BH: The parameters are $R_s=2$, $\gamma=1$, and $\Delta=0.25$.}\label{fig:transfer-functions}
\end{figure*}

Now we study the shadows of the two kinds of BHs surrounded by an optically and geometrically thin accretion disk on the equatorial plane of BHs, with an observer located at the north pole. Let us consider the simple case, where the emission originates from the accretion disk near BHs, and the emission intensity $I_{\nu}^{\rm em}$ depends only on the radial coordinate $r$. Here $\nu$ denotes the emission frequency in a static frame. It turns out that the observed intensity is related to the emission intensity by~\cite{Gralla:2019xty}
\begin{align}\label{eq:Iobs}
 I_{\rm obs}(b)&=\left.\sum_m f(r)^2I_{\rm em}(r)\right|_{r=r_m(b)},
\end{align}
where $I_{\rm em}(r):=\int I^{\rm em}_\nu(r){\rm d}\nu$ is the integrated intensity, and $r_m(b)$ ($m=1,2,3,\cdots$) is the so-called transfer function describing the radial position of the $m$th intersection of the light ray and the accretion disk outside the horizon at $\phi=\frac{2m-1}{2}\pi$. Here the absorption and reflection of light by the accretion disk are neglected for simplify. For the case of $m > 3$, it turns out that the contributions from the corresponding photon rings to the total luminosity can be ignored. The first three transfer functions $r_m(b)$ ($m=1,2,3$) can be expressed as
\begin{align}
 r_1(b)=\frac{1}{u\left(\frac{\pi}{2},b\right)}, \qquad b\in(b_1^-,\infty),\\
 r_2(b)=\frac{1}{u\left(\frac{3\pi}{2},b\right)}, \qquad b\in(b_2^-,b_2^+),\\
 r_3(b)=\frac{1}{u\left(\frac{5\pi}{2},b\right)}, \qquad b\in(b_3^-,b_3^+),
\end{align}
where $u(\phi,b)$ denotes the solution to the orbit equation \eqref{eq:orbit-eq-u-phi}. The first three transfer functions for Schwarzschild spacetime and the effective spacetime are plotted in Fig.~\ref{fig:transfer-functions}.

According to Eq.~\eqref{eq:Iobs}, the local brightness can potentially reach extremely high values as a result of the superposition of contributions from all the intersections. However, as emphasized in Ref.~\cite{Gralla:2019xty}, the detector should measure the average brightness, which is proportional to the flux detected by the detector. As shown in Fig.~\ref{fig:transfer-functions}, the second and third transfer functions for the quantum corrected BH with $k=0$ only have support within the ``photon ring'' regions $b\in (4.85475, 6.12553)$ and $b\in (5.07672, 5.13117)$ respectively.  Thus, the support of the higher-order transfer function is dramatically narrower than that of the lower-order one, since the light rays would converge due to the effect of the strong gravitational field, and hence those interacting with the accretion disk more times would converge towards a smaller region. Consequently, the images of the disk resulting from the contributions of $m>1$ are highly demagnified, with demagnification factors determined by the slope ${\rm d}r_m/{\rm d}b$. For the case of $m=3$, the numerical result yields ${\rm d}r_3/{\rm d}b\approx581.908$, indicating that the tertiary image contributes  around 1/580 of total flux. Since the images for $m>3$ are even greater demagnified compared to $m=3$, their contribution to the total flux is negligible. Moreover, as shown in Ref.~\cite{Gralla:2019xty}, the negligible contribution of the photon ring to the total flux can be analytically understood by examining the total azimuth angle $\phi$ given in Eqs.~\eqref{eq:totalangle1} and \eqref{eq:totalangle2}, with the corresponding numerical results displayed in Fig.~\ref{fig:trajectory}. By applying the method of matched asymptotic expansion (see Ref.~\cite{Zhang:2023okw} and references therein), it can be analytically calculated that $\phi(b)$ diverges logarithmically as $b$ approaches $b_c$, i.e., $\phi\sim -C \log|b-b_c|$, where $C$ is a positive constant depending on the sign of $b-b_c$. This logarithmic divergence is consistent with the divergent behavior in the Schwarzschild spacetime. Therefore, following the analysis presented in Ref.~\cite{Gralla:2019xty}, it can be concluded that the contributions of photon rings with $m>3$ to the total flux decay exponentially.

%================fig================
%===================================

\begin{figure*}
\includegraphics[scale=0.34]{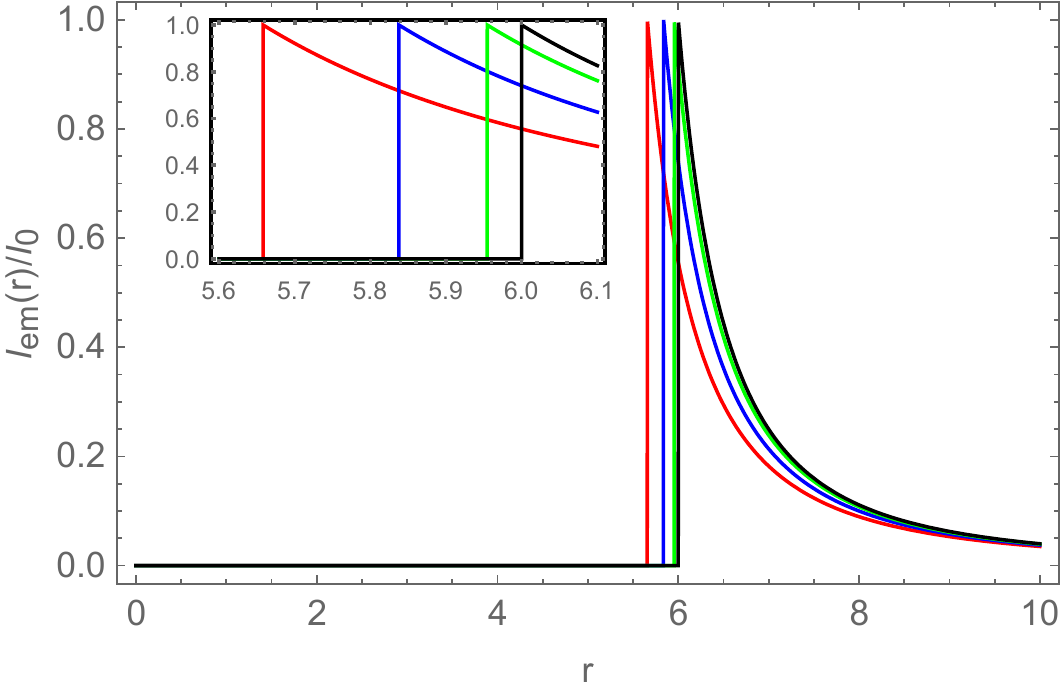}\quad
\includegraphics[scale=0.34]{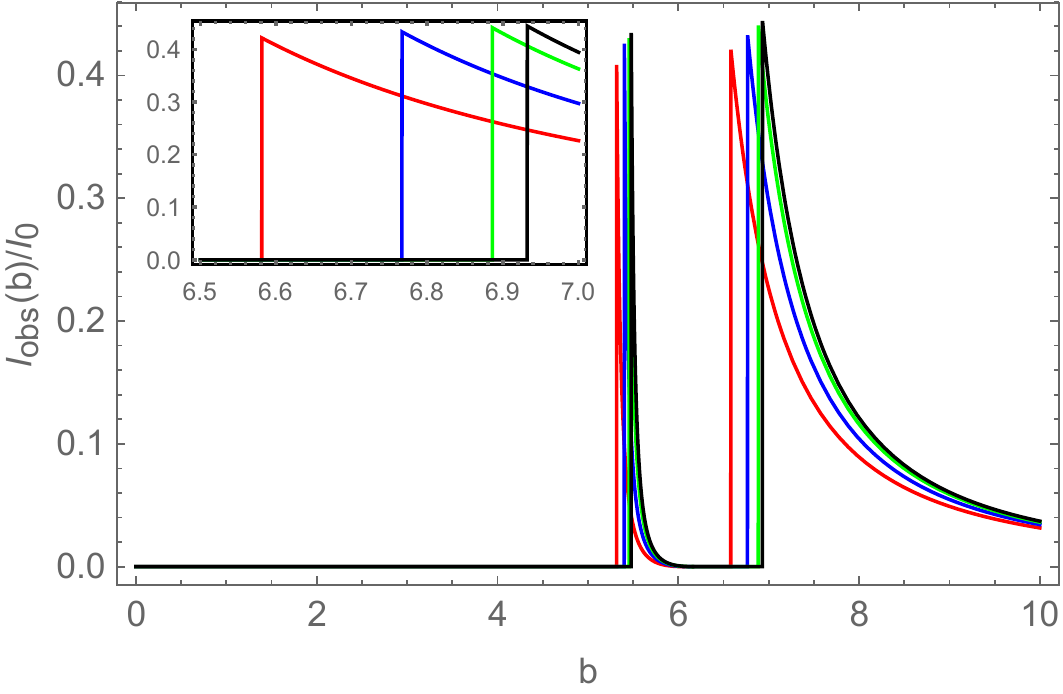}\quad
\includegraphics[scale=0.37]{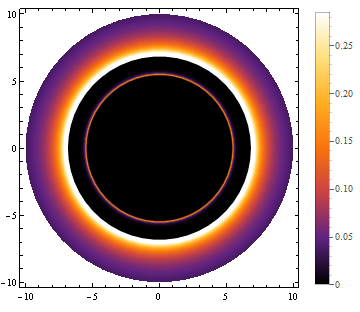}
\includegraphics[scale=0.34]{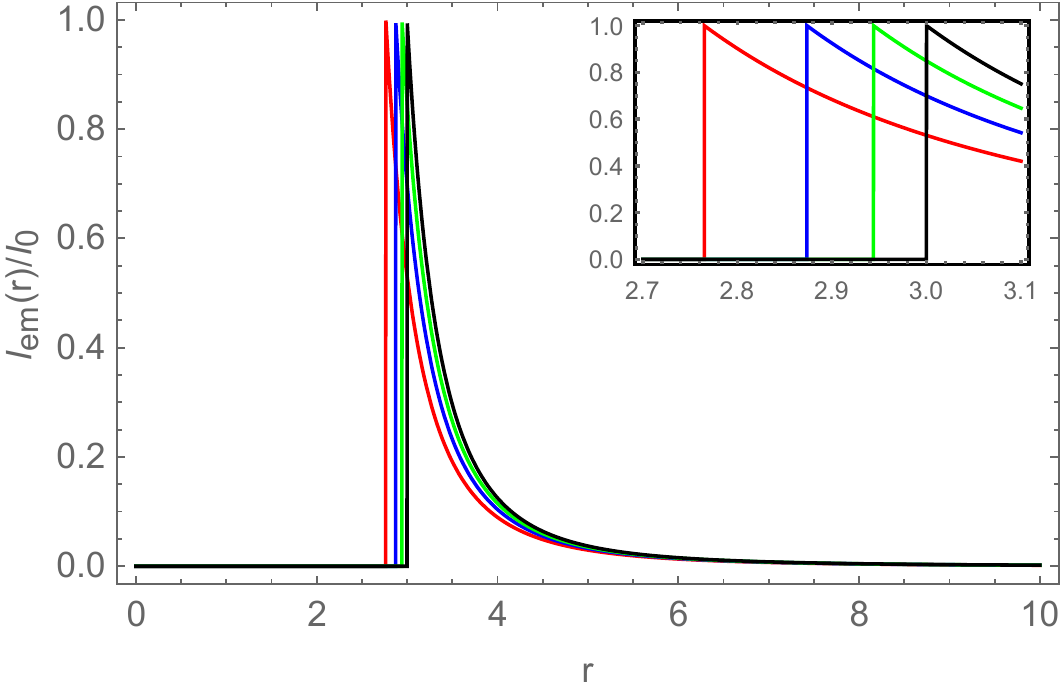}\quad
\includegraphics[scale=0.34]{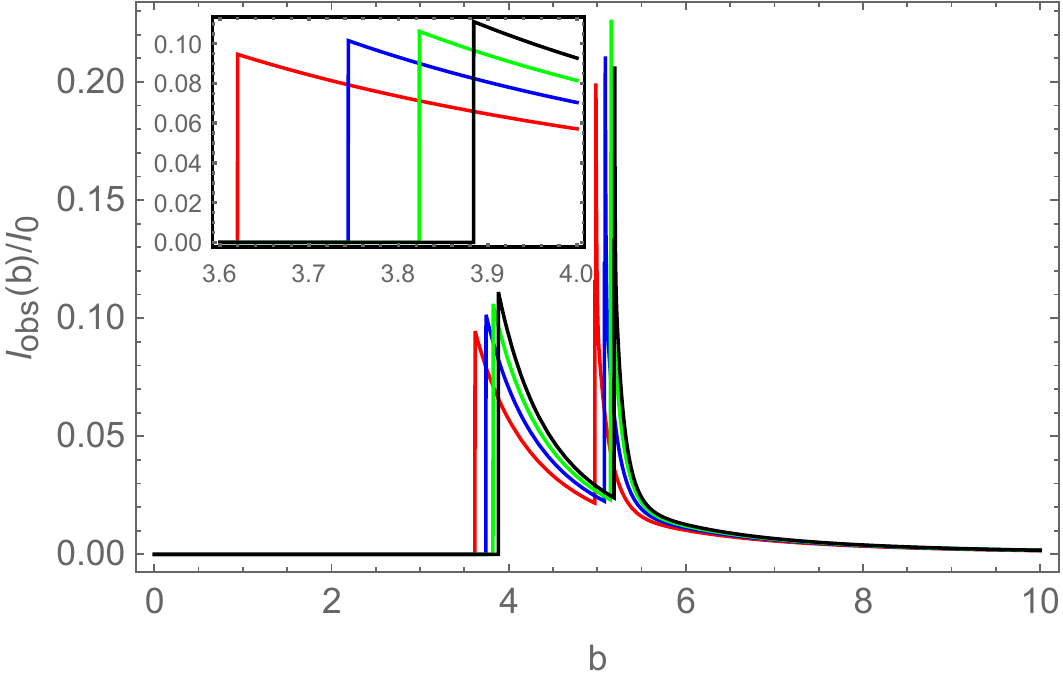}\quad
\includegraphics[scale=0.37]{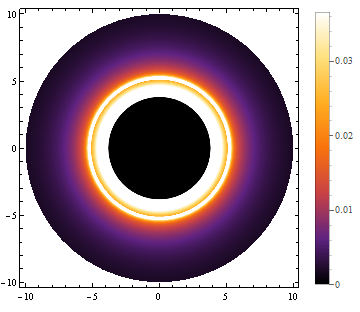}
\includegraphics[scale=0.34]{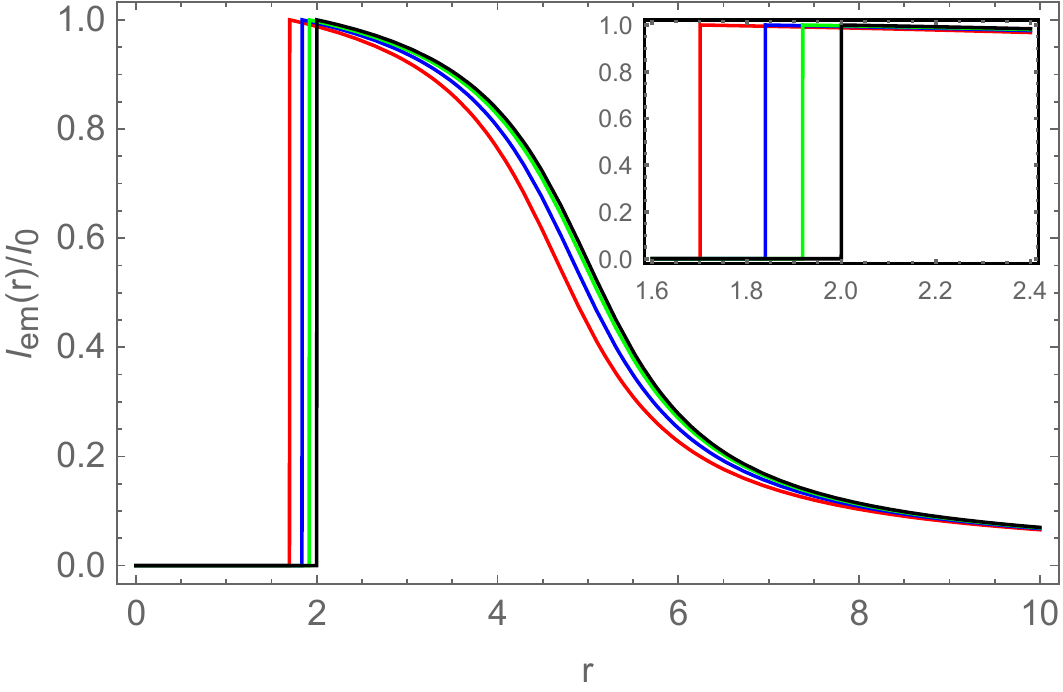}\quad
\includegraphics[scale=0.34]{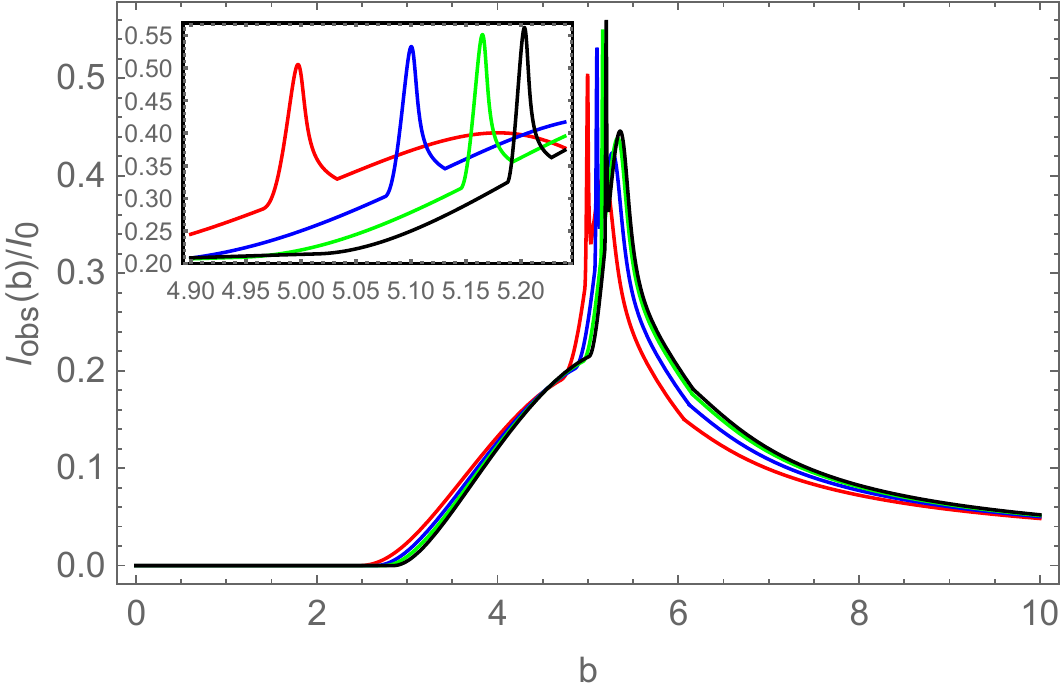}\quad
\includegraphics[scale=0.37]{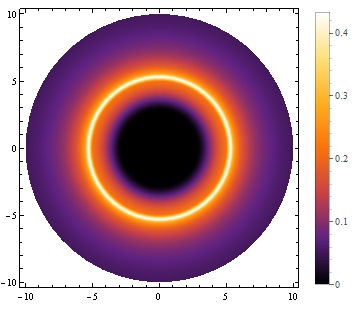}
\caption{The observational appearances of the thin disk near the BHs with the three different profiles: In each row, the first two panels show the emission intensity $I_{\rm em}/I_0$ and observational intensity $I_{\rm obs}/I_0$, normalized to the maximum value $I_0$, of a thin disk near the quantum-corrected BHs, corresponding to $k=-1$ (red), $k=0$ (blue) and $k=+1$ (green), compared to those of the Schwarzschild BH (black), and the third panel depicts the density plot of $I_{\rm obs}/I_0$ of a thin disk near the quantum-corrected BH with $k=0$. The parameters are $R_s=2$, $\gamma=1$, $\Delta=0.25$, and $\tilde{r}_0=0.5$.}\label{fig:shadow}
\end{figure*}

To study the observational appearance of emission, one needs to specify the intensity of emission $I_{\rm em}$. Now let us consider the following three specific intensities of emission,
\begin{align}
 I_{\rm em}(r)&:=
 \begin{cases}
  I_0\left[\frac{1}{r-(r_{\rm isco}-1)}\right]^2, &\hspace{1.1cm}  r>r_{\rm isco}\\
  0,&\hspace{1.1cm} r \leqslant r_{\rm isco}
 \end{cases},\label{eq:intensity-1}\\
 I_{\rm em}(r)&:=
 \begin{cases}
  I_0\left[\frac{1}{r-(r_{\rm ph}-1)}\right]^3, &\hspace{1.15cm}  r>r_{\rm ph}\\
  0,&\hspace{1.15cm} r\leqslant r_{\rm ph}
 \end{cases},\label{eq:intensity-2}\\
 I_{\rm em}(r)&:=
 \begin{cases}
  I_0\frac{\frac{\pi}{2}-\arctan[r-(r_{\rm isco}-1)]}{\frac{\pi}{2}-\arctan[r_h-(r_{\rm isco}-1)]}, &\quad r>r_h\\
  0,&\quad r\leqslant r_h
 \end{cases},\label{eq:intensity-3}
\end{align}
where the emission intensities are peaked at $r_{\rm isco}$, $r_{\rm ph}$ and $r_h$, respectively, and decay sharply for the first two cases and decay slowly for the last case. Here $I_0$ denote the maximum value of the emitted intensities. It is easy to see from Table \ref{table:1} that the difference of the three values of $r_{\rm isco}$, $r_{\rm ph}$ and $r_h$ between the Schwarzschild and effective spacetimes are very tiny, resulting in almost the same observational appearance of emission originating near BHs. The observational appearances of the thin disk near the BHs with these three different profiles are shown in Fig.~\ref{fig:shadow}. For comparison, the left and middle panels in each row of Fig.~\ref{fig:shadow} depict the emission intensities and observational intensities for the Schwarzschild BH and its quantum corrections. We observe that for the fixed parameters $R_s=2$, $\gamma=1$, $\Delta=0.25$, and $\tilde{r}_0=0.5$, the quantum correction always shrinks the radius of shadows.
%================fig================
%===================================

\subsection{Quasinormal modes}\label{sec-III-B}

QNMs are characterized by some complex frequencies of the linear perturbations around BH solutions (see~\cite{Kokkotas:1999bd,Konoplya:2011qq} for review). They describe the response of a BH to external perturbations. It is widely believed that the study of QNMs plays important roles not only in analyzing the stability of BHs, but also in understanding gravitational wave signals. The modes consist of the oscillation frequency (the real part) and the decay width (the imaginary part). In 1957, Regge and Wheeler studied the perturbation of Schwarzschild BH for the first time~\cite{Regge:1957td}. Since then, the QNMs of various BHs have been studied~\cite{Zerilli:1970se,Chandrasekhar:1975zza,Gundlach:1993tp,Schutz:1985km,Iyer:1986np,Wang:2000gsa,Li:2015mqa,Zou:2017juz,Zhang:2020sjh,Qian:2020wbv,Liu:2020evp,Liu:2021fzr,Wang:2021upj,Wang:2021uix}. Recently the perturbations of some LQBHs have also been studied~\cite{Dreyer:2002vy,Santos:2015gja,Cruz:2015bcj,Anacleto:2020zhp,Liu:2020ola,Bouhmadi-Lopez:2020oia,Cruz:2020emz,Daghigh:2020fmw,Santos:2021wsw,Liu:2021djf,del-Corral:2022kbk,Momennia:2022tug}. In this section, we will calculate the QNMs of the quantum-corrected BHs described in Sec.~\ref{sec-II} under certain perturbations. 

Considering the spherical symmetry of the spacetime, the perturbation field $\Psi$ can be expressed as
\begin{align}
  \Psi(t,r,\theta,\phi)=Y(\theta,\phi)\frac{\psi(t,r)}{r},
\end{align}
where $Y(\theta,\phi)$ denotes the spherical harmonics. It turns out that the equation of motion for the perturbation field $\Psi$ can be uniformly written in the following Schr\"odinger-like wave equation~\cite{Konoplya:2011qq}:
\begin{align}\label{eq:QNM-time-domain-eq}
 \frac{\partial^2\psi(t,r_*)}{\partial t^2}-\frac{\partial^2\psi(t,r_*)}{\partial r_*^2}+V(r_*)\psi(t,r_*)=0,
\end{align}
where $r_*$ is the tortoise coordinate as the solution of
\begin{align}\label{eq:tortoise-coordinate}
  {\rm d}r_*&=\frac{{\rm d}r}{f(r)},
\end{align}
which maps the region $(r_h,\infty)$ into the region $(-\infty,+\infty)$, and $V(r_*)$ denotes the effective potential with the form
\begin{align}\label{eq:QNM-V}
 V(r_*)&\equiv V(r(r_*))=f(r)\left[\frac{l(l+1)}{r^2}+\frac{1-s^2}{r}\frac{{\rm d}f(r)}{{\rm d}r}\right],
\end{align}
here $l$ is the multipole quantum number, and $s$ denotes the spin of the perturbation field with values $0$, $1$ and $2$ for the scalar perturbation, vector perturbation, and axial perturbation, respectively.

Assuming that $\psi(t,r_*)$ can be split into
\begin{align}
 \psi(t,r_*)=e^{-{\rm i}\omega t}\varphi(r_*),
\end{align}
Eq.~\eqref{eq:QNM-time-domain-eq} reduces to the time-independent wave equation
\begin{align}\label{eq:QNM-frequency-domain-eq}
 \frac{{\rm d}^2\varphi(r_*)}{{\rm d}r_*^2}+\left[\omega^2-V(r_*)\right]\varphi(r_*)=0.
\end{align}
The complex frequency $\omega$ can be obtained by solving Eq.~\eqref{eq:QNM-frequency-domain-eq} under appropriate boundary conditions for the wave at event horizon ($r_*=-\infty$), and infinity ($r_*=+\infty$).

\begin{figure*}
\includegraphics[scale=0.8]{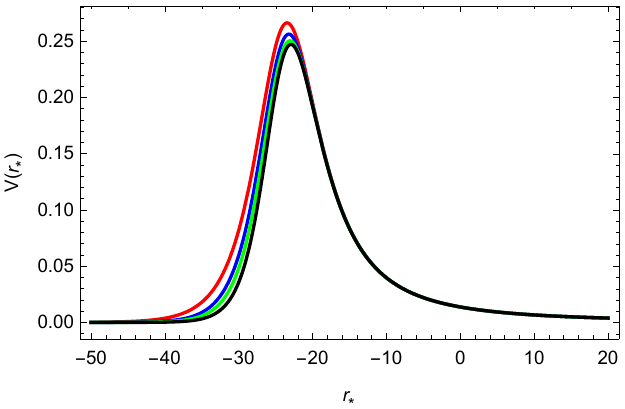}\qquad
\includegraphics[scale=0.8]{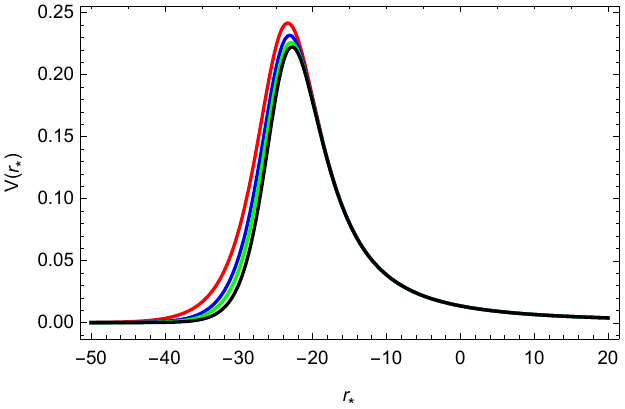}
\caption{Behaviors of the effective potential $V(r_*)$ for the scalar perturbation (left panel) and vector perturbation (right panel) of $l=2$, corresponding to the Schwarzschild BH (black) and the quantum-corrected BHs corresponding to $k=-1$ (red), $k=0$ (blue) and $k=+1$ (green): The parameters are $R_s=2$, $\gamma=1$, $\Delta=0.25$, and $\tilde{r}_0=0.5$. The profiles are obtained from Eq.~\eqref{eq:QNM-V} by inserting the solution to Eq.~\eqref{eq:tortoise-coordinate} with the initial data $r(r_*=0)=10R_s$.}\label{fig:V-rstar}
\end{figure*}

We now calculate the QNM frequencies $\omega$ under the scalar and vector perturbations in the case of $l=2$, starting from the time-independent wave equation \eqref{eq:QNM-frequency-domain-eq} and the time evolution wave equation \eqref{eq:QNM-time-domain-eq}, respectively. Notice that both Eq.~\eqref{eq:QNM-frequency-domain-eq} and Eq.~\eqref{eq:QNM-time-domain-eq} are completely determined by the effective potential $V(r_*)$. To obtain the precise values of $V$ at $r_*$, we integrate numerically Eq.~\eqref{eq:tortoise-coordinate} under suitable initial data, e.g., $r(r_*=0)=10R_s$, and inserting the result into the right-hand side of Eq.~\eqref{eq:QNM-V}. The effective potentials of the scalar and vector perturbations outside the classical and quantum-corrected BHs are plotted in Fig.~\ref{fig:V-rstar}. As it can be seen, for both perturbations, the effective potential for the effective theory always takes the maximum at each point $r_*$, and the effective potentials in the effective theory have the form of a barrier and take constant values at the event horizon and spatial infinity. Hence, it is convenient to impose the condition such that the wave is purely incoming at the horizon, while it is purely outgoing at spatial infinity, i.e., \cite{Schutz:1985km}
\begin{align}
 \varphi(r_*)&\sim 
 \begin{cases}
  e^{-{\rm i}\omega r_*}, \qquad r_*\rightarrow-\infty\\
  e^{+{\rm i}\omega r_*}, \qquad r_*\rightarrow+\infty
 \end{cases}.
\end{align}
Moreover, negative imaginary part of $\omega$ indicates that $\psi$ is damped and thus is stable, while positive imaginary part means an instability.
%================fig================
%===================================

%================table================
\begin{table*}
\caption{The QNMs $\omega$ of the scalar and vector perturbations obtained by the WKB method up to 6th order for the Schwarzschild BH (Sch) and the quantum-corrected BHs corresponding to $k=-1$ (QC-I), $k=0$ (QC-II) and $k=+1$ (QC-III), with $R_s=2$, $\gamma=1$, $\Delta=0.25$, and $\tilde{r}_0=0.5$.}
%\scriptsize
\footnotesize
\begin{ruledtabular}
\begin{tabular}{cccccccccc}\label{table:WKB}
 &&&Scalar perturbation\\
  BHs &  1st-order  & 2nd-order   & 3rd-order  & 4th-order & 5th-order & 6th-order   \\
  \hline
  Sch & $0.506317 - 0.0961232 \,{\rm i}$ & $0.483977 - 0.10056 \,{\rm i}$ & $0.483211 - 0.0968049 \,{\rm i}$ & $0.483647 - 0.0967175 \,{\rm i}$ & $0.483656 -  0.0967632 \,{\rm i}$ & $0.483642 - 0.0967661 \,{\rm i}$\\
  QC-I & $0.523674 - 0.0892331 \,{\rm i}$ & $ 0.503676 - 0.0927761 \,{\rm i}$ & $ 0.502963 - 0.0888253 \,{\rm i}$ & $ 0.503759 - 0.088685 \,{\rm i}$ & $ 0.50379 - 0.0888621 \,{\rm i}$ & $ 0.503754 - 0.0888685 \,{\rm i}$\\
  QC-II & $0.514466 - 0.0917491 \,{\rm i}$ & $ 0.493705 - 0.0956073 \,{\rm i}$ & $ 0.492965 - 0.0917117 \,{\rm i}$ & $ 0.493701 - 0.0915751 \,{\rm i}$ & $ 0.493733 - 0.0917496 \,{\rm i}$ & $ 0.493691 - 0.0917576 \,{\rm i}$\\
  QC-III & $0.508967 - 0.0934035 \,{\rm i}$ & $ 0.48751 - 0.0975144 \,{\rm i}$ & $ 0.486741 - 0.0935928 \,{\rm i}$ & $ 0.487422 - 0.0934621 \,{\rm i}$ & $ 0.487452 - 0.0936213 \,{\rm i}$ & $ 0.487407 - 0.0936299 \,{\rm i}$\\
  \hline
    \hline
  &&&Vector perturbation\\
  BHs &  1st-order  & 2nd-order   & 3rd-order  & 4th-order & 5th-order & 6th-order   \\
  \hline
  Sch & $0.480754 - 0.0943536 \,{\rm i}$ & $0.457976 - 0.0990466 \,{\rm i}$ & $0.457131 - 0.0950652 \,{\rm i}$ & $0.457596 - 0.0949685 \,{\rm i}$ & $0.457605 - 0.0950088 \,{\rm i}$ & $0.457593 - 0.0950112 \,{\rm i}$\\
  QC-I & $0.499079 - 0.0873723 \,{\rm i}$ & $ 0.47887 - 0.0910596 \,{\rm i}$ & $ 0.478095 - 0.086892 \,{\rm i}$ & $ 0.478992 - 0.0867293 \,{\rm i}$ & $ 0.479032 - 0.0869498 \,{\rm i}$ & $ 0.47898 - 0.0869592 \,{\rm i}$\\
  QC-II & $0.489533 - 0.0899538 \,{\rm i}$ & $ 0.468486 - 0.0939949 \,{\rm i}$ & $ 0.467681 - 0.089894 \,{\rm i}$ & $ 0.468491 - 0.0897386 \,{\rm i}$ & $ 0.46853 - 0.0899428 \,{\rm i}$ & $ 0.468478 - 0.0899527 \,{\rm i}$\\
  QC-III & $0.483787 - 0.0916002 \,{\rm i}$ & $ 0.461997 - 0.0959203 \,{\rm i}$ & $ 0.461158 - 0.0917914 \,{\rm i}$ & $ 0.461896 - 0.0916448 \,{\rm i}$ & $ 0.46193 - 0.0918208 \,{\rm i}$ & $ 0.461883 - 0.0918303 \,{\rm i}$
\end{tabular}
\end{ruledtabular}
\end{table*}
%===================================

%================fig================
\begin{figure*}
\includegraphics[scale=0.5]{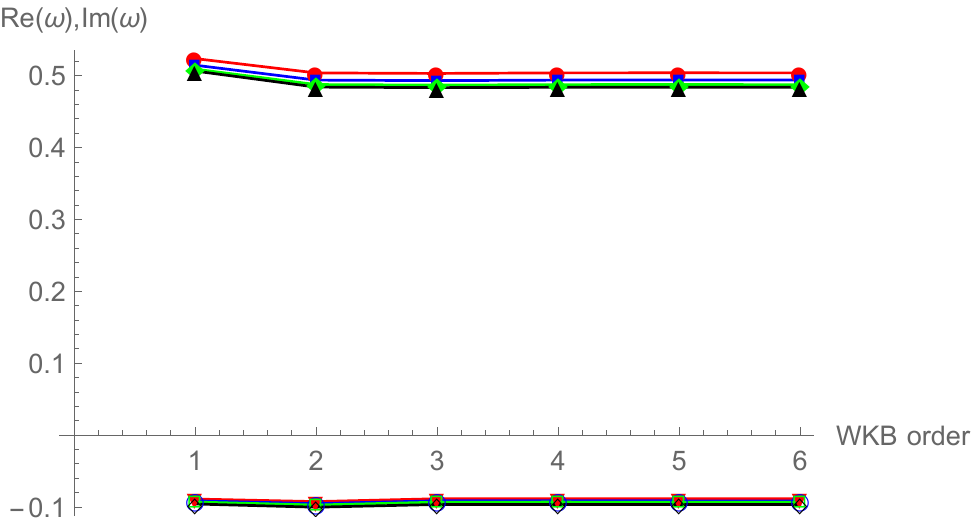}\quad
\includegraphics[scale=0.5]{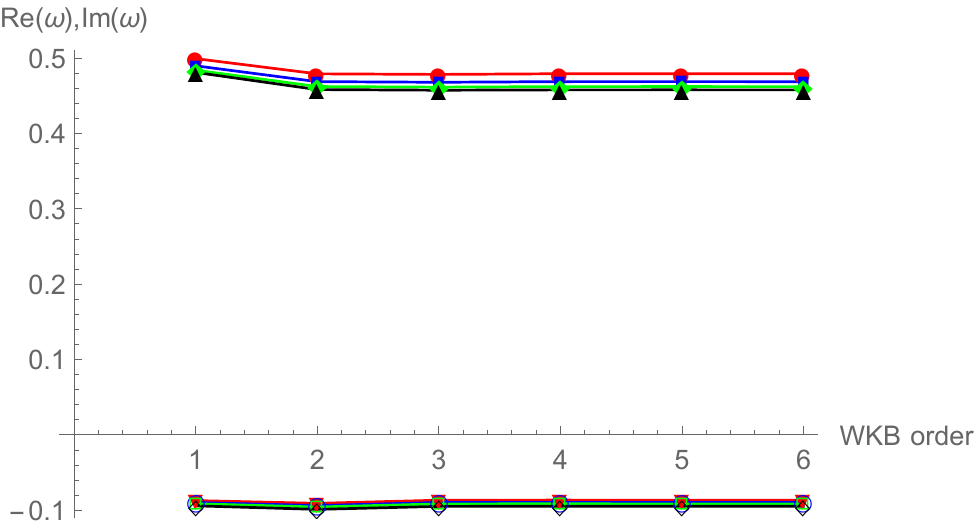}
\caption{The real (top) and imaginary (bottom) parts of the frequencies $\omega$ as a function of the order of the WKB method up to 6th order for the scalar (left panel) and vector (right panel) perturbations of $l=2$, corresponding to the Schwarzschild BH (black) and the quantum-corrected BHs corresponding to $k=-1$ (red), $k=0$ (blue) and $k=+1$ (green). The parameters are $R_s=2$, $\gamma=1$, $\Delta=0.25$, and $\tilde{r}_0=0.5$.}\label{fig:QNM-WKB}
\end{figure*}
%===================================

\begin{figure*}
\includegraphics[scale=0.6]{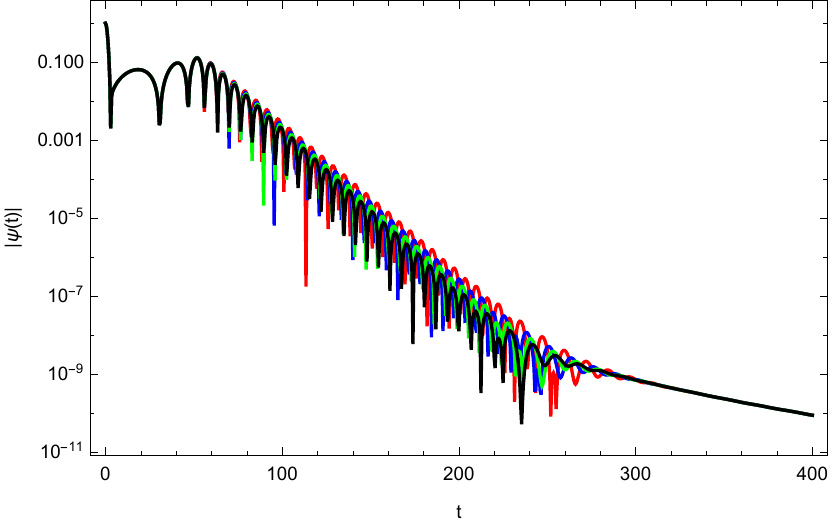}\qquad
\includegraphics[scale=0.6]{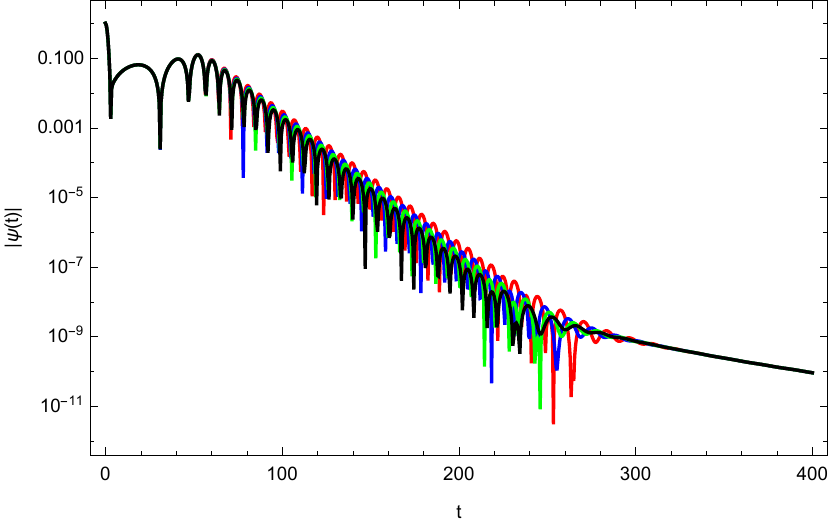}
\caption{The time domain profiles of the scalar (left panel) and vector (right panel) perturbations, corresponding to the Schwarzschild BH (black) and the quantum-corrected BHs corresponding to $k=-1$ (red), $k=0$ (blue) and $k=+1$ (green). The time domain profiles are observed at $r_*=0$ (or $r=10R_s$) under the initial Gaussian wavepacket \eqref{eq:gaussian-wavepacket} with $w_0=0$, $\sigma=1$, and the grid spacing $\delta=0.5$. The parameters are $R_s=2$, $\gamma=1$, $\Delta=0.25$, and $\tilde{r}_0=0.5$.}\label{fig:QNM-profile}
\end{figure*}

To solve $\omega$ from Eq.~\eqref{eq:QNM-frequency-domain-eq}, we adopt the WKB method, by treating it as the problem of scattering near the peak of the barrier potential in quantum mechanics. This method is initially utilized in~\cite{Schutz:1985km}, and is improved in~\cite{Konoplya:2003ii,Konoplya:2019hlu}. The resulting values for the fundamental QNM frequencies up to the 6th-order approximation are shown in Table \ref{table:WKB} as well as in Fig.~\ref{fig:QNM-WKB}. Also, one can employ the time domain integration method to calculate the evolution of $\psi(t,r_*)$ at a fixed point $r_*$ and obtain the time-domain profile, from which the frequency $\omega$ can be extracted by using the Prony method (See, e.g.,~\cite{Marple:1987,Berti:2007dg} for reference). To do this, we implement the time domain integration by employing the finite difference method. Introducing the light-like coordinates
\begin{align}
 u:=t-r_*,\qquad w:=t+r_*,
\end{align}
then Eq.~\eqref{eq:QNM-time-domain-eq} can be expressed in terms of $u$ and $w$ as
\begin{align}\label{eq:QNM-evolution-diff}
 -4\frac{\partial^2\psi(u,w)}{\partial u\partial w}=V\left(\frac{w-u}{2}\right)\psi(u,w).
\end{align}
To study numerically the differential equation \eqref{eq:QNM-evolution-diff}, one can discretize it on the $u$-$w$ null grid as~\cite{Gundlach:1993tp}
\begin{align}\label{eq:QNM-evolution-discrete}
 \psi_{i,j}=&\psi_{i,j-1}+\psi_{i-1,j}-\psi_{i-1,j-1}-\frac{\delta^2}{8}V_{i-1,j-1}\left(\psi_{i,j-1}+\psi_{i-1,j}\right)\notag\\
 &+O(\delta^4),
\end{align}
where $\delta$ denotes the overall grid scale factor, $\psi_{i,j}:=\psi(i\delta,j\delta)$ and $V_{i,j}:=V\left(\frac{(j-i)\delta}{2}\right)$. To implement the discretized evolution \eqref{eq:QNM-evolution-discrete}, one needs to specify certain initial data. It is shown that the QNMs depend neither on the initial data nor on the small $\delta$. Following Ref.~\cite{Gundlach:1993tp}, on the null boundary $u=0$, $\psi$ is specified as a Gaussian wavepacket
\begin{align}\label{eq:gaussian-wavepacket}
 \psi_{0,j}:=e^{-\frac{(j\delta-w_0)^2}{2\sigma^2}},
\end{align}
where $w_0$ and $\sigma$ are the median and width of the wavepacket, and on the null boundary $w=0$, $\psi$ is specified as a constant determined by $\psi_{0,0}$. Then one can calculate the values of $\psi_{i,j}$ on the whole grid from Eq.~\eqref{eq:QNM-evolution-discrete}. In this way, one can generate the ringdown waveform, namely, the time domain profile of perturbations by extracting the values of $\psi$ at constant $r_*$. To determine the quasinormal frequency corresponding to the profile, we use the Prony method to fit the profile data, and set $w_0=0$, $\sigma=1$, $\delta=0.5$. The ringdown waveforms $|\psi|$ of the classical and quantum-corrected BHs under the scalar and vector perturbations with $l=2$ are plotted in Fig.~\ref{fig:QNM-profile}. It shows that the quantum-corrected BHs and the Schwarzschild BH have almost identical waveforms. All the waveforms have the same power-law tail due to the fact that the effective potentials of these BHs have the same asymptotic behavior. We observe that the oscillation frequency for each of the quantum-corrected BHs is higher than that of the Schwarzschild one in each of the perturbations. Moreover, the damping rate for the Schwarzschild BH is the higher in each of the perturbations. These results are in agreement with those provided in Table \ref{table:WKB}. To further check the consistency between the generated ringdown waveforms in Fig.~\ref{fig:QNM-profile} and the QNMs obtained by the WKB method and presented in Table \ref{table:WKB}, the frequencies $\omega$ are also extracted by the Prony method from the ringdown waveforms in Fig.~\ref{fig:QNM-profile} and shown in Table \ref{table:Prony}. By comparison, we find that the data presented in Table \ref{table:WKB} agree quite well with those shown in Table \ref{table:Prony}. The fact that the imaginary frequencies always take negative values indicates that these oscillations will die off with time evolution, and thus the classical and quantum-corrected BHs are all stable against the scalar and vector perturbations.
%================fig================
%================table================
\begin{table}
\caption{The QNMs $\omega$ of the scalar and vector perturbations of $l=2$ extracted by the Prony method from the profile shown in Fig.~\ref{fig:QNM-profile} with $t\in[70,150]$ for the Schwarzschild BH (Sch) and the quantum-corrected BHs corresponding to $k=-1$ (QC-I), $k=0$ (QC-II) and $k=+1$ (QC-III): The parameters are $R_s=2$, $\gamma=1$, $\Delta=0.25$, and $\tilde{r}_0=0.5$.}
%\scriptsize
%\footnotesize
\begin{ruledtabular}
\begin{tabular}{cccccccccc}\label{table:Prony}
  BHs &  $\omega$ (scalar perturbation)  & $\omega$ (vector perturbation)     \\
  \hline
  Sch    & $0.484111  -0.0959327 \,{\rm i}$ & $0.458259 -0.0942558 \,{\rm i}$ \\
  QC-I   & $0.504153 -0.0879953 \,{\rm i}$ & $0.479610 -0.0861299 \,{\rm i}$ \\
  QC-II  & $0.494134 -0.0908975 \,{\rm i}$ & $0.469145 -0.0891647 \,{\rm i}$ \\
  QC-III & $0.487869 -0.0927798 \,{\rm i}$ & $0.462559 -0.0910634 \,{\rm i}$     
\end{tabular}
\end{ruledtabular}
\end{table}

\section{Summary}\label{sec-IV}

To understand how quantum corrections affect the observables of BHs, the shadows and QNMs of quantum-corrected BHs have also been studied in the previous sections. On one hand, we considered the shadows and appearances of BHs surrounded by the optically and geometrically thin accretion disk on the equatorial plane of BHs with the specified intensities of emission $I_{\rm em}$ by Eqs.~\eqref{eq:intensity-1}--\eqref{eq:intensity-3}. It has been shown in Sec.~\ref{sec-III-A} that the quantum correction always shrinks the radius of shadows. On the other hand, in Sec.~\ref{sec-III-B} we calculated the QNMs of the quantum-corrected BHs under the scalar and vector perturbations by two methods, the WKB approximation method and the time domain integration method. The results are compared with the Schwarzschild case and presented in Tables~\ref{table:WKB} and \ref{table:Prony} as well as Figs.~\ref{fig:QNM-WKB} and \ref{fig:QNM-profile}. The data presented in Table~\ref{table:WKB} agree quite well with those shown in Table \ref{table:Prony}. Moreover, we found that the quantum correction increases the real part of the QNMs and decreases the absolute value of the imaginary part. In particularly, the imaginary frequencies of QNMs always take negative values, and thus the quantum-corrected BHs are stable against the scalar and vector perturbations.

%-------------------------------------------------------------
\begin{acknowledgments}
This work is supported in part by NSFC Grants No.~12165005, No.~12275022 and No.~11961131013. C. Z. acknowledges the support by the Polish Narodowe Centrum Nauki, Grant No. 2018/30/Q/ST2/00811.
\end{acknowledgments}

%-------------------------------------------------------------
\begin{appendix}

\section{Junction conditions}\label{appendix}
In the exterior region ${\cal M}^+$, we assume that the metric is still stationary and can be expressed in Schwarzschild coordinates $(t,r,\theta,\phi)$ by the general form
\begin{align}\label{eq:line-element-exterior-general}
  {\rm d}s_+^2=-F(r){\rm d}t^2+L(r)^{-1}{\rm d}r^2+r^2{\rm d}\Omega^2.
\end{align}
For the special case of $f(r)=F(r)=L(r)$, Eq.~\eqref{eq:line-element-exterior-general} reduces to Eq.~\eqref{eq:line-element-exterior}. As seen from the inside, the induced line element on $\Sigma$ from the line element \eqref{eq:line-element-interior} of ${\cal M}^-$ reads
\begin{align}\label{eq:line-element-interior-induced}
  \left.{\rm d}s_-^2\right|_\Sigma=-{\rm d}\tau^2+a(\tau)^2\tilde{r}_0^2\,{\rm d}\Omega^2.
\end{align}
As seen from the outside, the surface $\Sigma$ can be described by the parametric equations $r=r(\tau)$ and $t=t(\tau)$. Thus, the induced line element on $\Sigma$ from the exterior line element \eqref{eq:line-element-exterior-general} is given by
\begin{align}\label{eq:line-element-exterior-induced}
  \left.{\rm d}s_+^2\right|_\Sigma=-\left(F\dot{t}^2-L^{-1}\dot{r}^2\right){\rm d}\tau^2+r(\tau)^2{\rm d}\Omega^2,
\end{align}
where a dot over a letter denotes its derivative with respect to $\tau$.

To match the exterior region ${\cal M}^+$ with the interior region ${\cal M}^-$ along the boundary surface $\Sigma$ so that $\Sigma$ forms a unique surface in the entire spacetime ${\cal M}$, one needs to impose suitable boundary conditions (junction conditions). In the classical theory, the Darmois-Israel junction conditions~\cite{Darmois:1927,Israel:1966rt,Poisson:2004bk} require that the first and the second fundamental forms on the two sides of the boundary surface $\Sigma$ equal to each other, respectively. In our case, the match of the first fundamental form yields
\begin{align}
 &1=F\dot{t}^2-L^{-1}\dot{r}^2\quad\Leftrightarrow\quad F\dot{t}=\sqrt{\dot{r}^2FL^{-1}+F}\equiv\beta(r,\dot{r}),\label{eq:match-qtt}\\
 &a(\tau)\tilde{r}_0=r(\tau)\label{eq:match-qrr}.
\end{align}
To calculate the extrinsic curvature (the second fundamental form) of $\Sigma$, we chose the direction of its normal towards ${\cal M}^+$. Then the components of extrinsic curvature as seen from ${\cal M}^-$ read
\begin{align}\label{eq:curvature-interior}
K^-_{\tau\tau}=0,\quad
K^-_{\theta\theta}=\frac{K^-_{\phi\phi}}{\sin^2\theta}&=a(\tau)\tilde{r}_0\sqrt{1-k\tilde{r}_0^2}.
\end{align}
By using Eq.~\eqref{eq:match-qtt}, the components of extrinsic curvature as seen from ${\cal M}^+$ can be calculated as
\begin{align}\label{eq:curvature-exterior}
K^+_{\tau\tau}=-\frac{\dot{\beta}}{\dot{r}}\sqrt{F^{-1}L},\quad
K^+_{\theta\theta}=\frac{K^+_{\phi\phi}}{\sin^2\theta}&=r\beta\sqrt{F^{-1}L}.
\end{align}
Hence matching the extrinsic curvatures on the two sides of $\Sigma$ leads to
\begin{align}
0&=-\frac{\dot{\beta}}{\dot{r}}\sqrt{F^{-1}L},\label{eq:match-ktautau}\\
a(\tau)\tilde{r}_0\sqrt{1-k\tilde{r}_0^2}&=r\beta\sqrt{F^{-1}L}\label{eq:match-kphiphi}.
\end{align}
Up to now, we have obtained the junction conditions determined completely by Eqs.~\eqref{eq:match-qtt}, \eqref{eq:match-qrr}, \eqref{eq:match-ktautau}, and \eqref{eq:match-kphiphi}. To satisfy Eq.~\eqref{eq:match-ktautau}, it is sufficient to require that the integral curve of $\partial/\partial\tau$ is geodesic in $\mathcal M^+$. Since $\partial/\partial t$ is a Killing vector field in $\mathcal M^+$,  $\beta$ is just  the conserved  energy along the geodesic tangent to $\partial/\partial \tau$, i.e.,
\begin{align}
E=-g^+_{ab}\left(\frac{\partial}{\partial t}\right)^a\left(\frac{\partial}{\partial \tau}\right)^b=F\dot{t}=\beta,
\end{align}
which leads to $\dot{\beta}=0$. Then, combing Eq.~\eqref{eq:match-kphiphi} with Eq.~\eqref{eq:match-qrr}, we get
\begin{align}\label{eq:beta-FL}
\beta=\sqrt{1-k\tilde{r}_0^2}\sqrt{FL^{-1}}.
\end{align}
Combing Eq.~\eqref{eq:match-qtt} with Eq.~\eqref{eq:beta-FL} and annihilating $\beta$, we have
\begin{align}\label{eq:rdot-L}
\dot{r}^2=\left(1-k\tilde{r}_0^2\right)-L.
\end{align}
Taking the derivative of  Eq.~\eqref{eq:match-qrr} with respect to $\tau$ and using Eq.~\eqref{eq:rdot-L}, we arrive at
\begin{align}\label{eq:L-H}
L=\left(1-k\tilde{r}_0^2\right)-H^2r^2.
\end{align}
Thus, by matching ${\cal M}^+$ to ${\cal M}^-$ along $\Sigma$ generated by geodesics, the junction conditions lead to the three equations \eqref{eq:match-qrr}, \eqref{eq:beta-FL}, and \eqref{eq:L-H}. Equation \eqref{eq:match-qrr} gives a relation between the two coordinates on $\Sigma$, while Eqs.~\eqref{eq:beta-FL} and \eqref{eq:L-H} determine a dynamical relation between the two regions.

In classical case, by inputting the dynamical equation \eqref{eq:classical-Friedmann-eq} in ${\cal{M}^-}$ into Eq.~\eqref{eq:L-H} and using Eq.~\eqref{eq:match-qrr}, we find
\begin{align}\label{eq:G-r-classical-1}
L=\left(1-k\tilde{r}_0^2\right)+k\frac{r^2}{a^2}-\frac{8\pi G}{3}\rho r^2=1-\frac{8\pi G}{3}\rho r^2=1-\frac{R_s}{r}.
\end{align}
Combing Eq.~\eqref{eq:G-r-classical-1} with \eqref{eq:beta-FL} yields
\begin{align}
F=\frac{\beta^2}{1-k\tilde{r}_0^2}\left(1-\frac{R_s}{r}\right).
\end{align}
Hence, the exterior metric \eqref{eq:line-element-exterior-general} in ${\cal M}^-$ can be generated as
\begin{align}\label{eq:line-element-exterior-classical-org}
  {\rm d}s_+^2&=-\frac{\beta^2}{1-k\tilde{r}_0^2}\left(1-\frac{R_s}{r}\right){\rm d}t^2+\left(1-\frac{R_s}{r}\right)^{-1}{\rm d}r^2+r^2{\rm d}\Omega^2.
\end{align}
Absorbing $\beta/\sqrt{1-k\tilde{r}_0^2}$ into $t$ in Eq.~\eqref{eq:line-element-exterior-classical-org} yields the Schwarzschild metric \eqref{eq:line-element-exterior-classical}.

\end{appendix}

%===============================================================================================

%===============================================================================================


\begin{thebibliography}{120}

\bibitem{Penrose:1964wq}
  R.~Penrose,
  {Gravitational collapse and space-time singularities}.
  Phys. Rev. Lett. \textbf{14}, 57 (1965).
  {\url{https://doi.org/10.1103/PhysRevLett.14.57}}

\bibitem{Hawking:1970zqf}
  S.W. Hawking, R.~Penrose,
  {The Singularities of gravitational collapse and cosmology}.
  Proc. Roy. Soc. Lond. A \textbf{314}, 529 (1970).
  {\url{https://doi.org/10.1098/rspa.1970.0021}}

\bibitem{Rovelli:2004tv}
  C.~Rovelli, \emph{{Quantum Gravity}}
  (Cambridge University Press, Cambridge, England, 2004)

\bibitem{Thiemann:2007pyv}
  T.~Thiemann, \emph{{Modern Canonical Quantum General Relativity}}
  (Cambridge University Press, Cambridge, England, 2007).
  {\url{https://doi.org/10.1017/CBO9780511755682}}

\bibitem{Thiemann:2002nj}
  T.~Thiemann,
  {Lectures on loop quantum gravity}.
  Lect. Notes Phys. \textbf{631}, 41 (2003).
  {\url{https://doi.org/10.1007/978-3-540-45230-0_3}}.
  {\href{https://arxiv.org/abs/gr-qc/0210094}{{arXiv:gr-qc/0210094}}}

\bibitem{Ashtekar:2004eh}
  A.~Ashtekar, J.~Lewandowski,
  {Background independent quantum gravity: A status report}.
  Class. Quant. Grav. \textbf{21}, R53 (2004).
  {\url{https://doi.org/10.1088/0264-9381/21/15/R01}}.
  {\href{https://arxiv.org/abs/gr-qc/0404018}{{arXiv:gr-qc/0404018}}}

\bibitem{Han:2005km}
  M.~Han, Y.~Ma, W.~Huang,
  {Fundamental structure of loop quantum gravity}.
  Int. J. Mod. Phys. D \textbf{16}, 1397 (2007).
  {\url{https://doi.org/10.1142/S0218271807010894}}.
  {\href{https://arxiv.org/abs/gr-qc/0509064}{{arXiv:gr-qc/0509064}}}

\bibitem{Giesel:2012ws}
  K.~Giesel, H.~Sahlmann,
  {From classical to quantum gravity: Introduction to loop quantum gravity}.
  Proc. Sci., \textbf{QGQGS2011}, 002 (2011).
  {\url{https://doi.org/10.22323/1.140.0002}}.
  {\href{https://arxiv.org/abs/1203.2733}{{arXiv:1203.2733}}}

\bibitem{Rovelli:2011eq}
  C.~Rovelli,
  {Zakopane lectures on loop gravity}.
  Proc. Sci., \textbf{QGQGS2011}, 003 (2011).
  {\url{https://doi.org/10.22323/1.140.0003}}.
  {\href{https://arxiv.org/abs/1102.3660}{{arXiv:1102.3660}}}

\bibitem{Perez:2012wv}
  A.~Perez,
  {The spin-foam approach to quantum gravity}.
  Living Rev. Relativity \textbf{16}, 3 (2013).
  {\url{https://doi.org/10.12942/lrr-2013-3}}.
  {\href{https://arxiv.org/abs/1205.2019}{{arXiv:1205.2019}}}

\bibitem{Rovelli:1994ge}
  C.~Rovelli, L.~Smolin,
  {Discreteness of area and volume in quantum gravity}.
  Nucl. Phys. B \textbf{442}, 593 (1995).
  {\url{https://doi.org/10.1016/0550-3213(95)00150-Q}}.
  {\href{https://arxiv.org/abs/gr-qc/9411005}{{arXiv:gr-qc/9411005}}}

\bibitem{Ashtekar:1996eg}
  A.~Ashtekar, J.~Lewandowski,
  {Quantum theory of geometry: I. Area operators}.
  Class. Quant. Grav. \textbf{14}, A55 (1997).
  {\url{https://doi.org/10.1088/0264-9381/14/1A/006}}.
  {\href{https://arxiv.org/abs/gr-qc/9602046}{{arXiv:gr-qc/9602046}}}

\bibitem{Ashtekar:1997fb}
  A.~Ashtekar, J.~Lewandowski,
  {Quantum theory of geometry II: Volume operators}.
  Adv. Theor. Math. Phys. \textbf{1}, 388 (1997).
  {\url{https://doi.org/10.4310/ATMP.1997.v1.n2.a8}}.
  {\href{https://arxiv.org/abs/gr-qc/9711031}{{arXiv:gr-qc/9711031}}}

\bibitem{Yang:2016kia}
  J.~Yang, Y.~Ma,
  {New volume and inverse volume operators for loop quantum gravity}.
  Phys. Rev. D \textbf{94}, 044003 (2016).
  {\url{https://doi.org/10.1103/PhysRevD.94.044003}}.
  {\href{https://arxiv.org/abs/1602.08688}{{arXiv:1602.08688}}}

\bibitem{Thiemann:1996at}
  T.~Thiemann,
  {A length operator for canonical quantum gravity}.
  J. Math. Phys. \textbf{39}, 3372 (1998).
  {\url{https://doi.org/10.1063/1.532445}}.
  {\href{https://arxiv.org/abs/gr-qc/9606092}{{arXiv:gr-qc/9606092}}}

\bibitem{Ma:2010fy}
  Y.~Ma, C.~Soo, J.~Yang,
  {New length operator for loop quantum gravity}.
  Phys. Rev. D \textbf{81}, 124026 (2010).
  {\url{https://doi.org/10.1103/PhysRevD.81.124026}}.
  {\href{https://arxiv.org/abs/1004.1063}{{arXiv:1004.1063}}}

\bibitem{Ashtekar:1997yu}
  A.~Ashtekar, J.~Baez, A.~Corichi, K.~Krasnov,
  {Quantum Geometry and Black Hole Entropy}.
  Phys. Rev. Lett. \textbf{80}, 904 (1998).
  {\url{https://doi.org/10.1103/PhysRevLett.80.904}}.
  {\href{https://arxiv.org/abs/gr-qc/9710007}{{arXiv:gr-qc/9710007}}}

\bibitem{Song:2020arr}
  S.~Song, H.~Li, Y.~Ma, C.~Zhang,
  {Entropy of black holes with arbitrary shapes in loop quantum gravity}.
  Sci. China Phys. Mech. Astron. \textbf{64}, 120411 (2021).
  {\url{https://doi.org/10.1007/s11433-021-1770-3}}.
  {\href{https://arxiv.org/abs/2002.08869}{{arXiv:2002.08869}}}

\bibitem{Thiemann:1996aw}
  T.~Thiemann,
  {Quantum spin dynamics (QSD)}.
  Class. Quant. Grav. \textbf{15}, 839 (1998).
  {\url{https://doi.org/10.1088/0264-9381/15/4/011}}.
  {\href{https://arxiv.org/abs/gr-qc/9606089}{{arXiv:gr-qc/9606089}}}

\bibitem{Thiemann:1997rt}
  T.~Thiemann,
  {Quantum spin dynamics (QSD): V. Quantum gravity as the natural regulator of
  matter quantum field theories}.
  Class. Quant. Grav. \textbf{15}, 1281 (1998).
  {\url{https://doi.org/10.1088/0264-9381/15/5/012}}.
  {\href{https://arxiv.org/abs/gr-qc/9705019}{{arXiv:gr-qc/9705019}}}

\bibitem{Yang:2015zda}
  J.~Yang, Y.~Ma,
  {New Hamiltonian constraint operator for loop quantum gravity}.
  Phys. Lett. B \textbf{751}, 343 (2015).
  {\url{https://doi.org/10.1016/j.physletb.2015.10.062}}.
  {\href{https://arxiv.org/abs/1507.00986}{{arXiv:1507.00986}}}

\bibitem{Alesci:2015wla}
  E.~Alesci, M.~Assanioussi, J.~Lewandowski, I.~M{\"a}kinen,
  {Hamiltonian operator for loop quantum gravity coupled to a scalar field}.
  Phys. Rev. D \textbf{91}, 124067 (2015).
  {\url{https://doi.org/10.1103/PhysRevD.91.124067}}.
  {\href{https://arxiv.org/abs/1504.02068}{{arXiv:1504.02068}}}

\bibitem{Zhang:2018wbc}
  C.~Zhang, J.~Lewandowski, Y.~Ma,
  {Towards the self-adjointness of a Hamiltonian operator in loop quantum
  gravity}.
  Phys. Rev. D \textbf{98}, 086014 (2018).
  {\url{https://doi.org/10.1103/PhysRevD.98.086014}}.
  {\href{https://arxiv.org/abs/1805.08644}{{arXiv:1805.08644}}}

\bibitem{Zhang:2019dgi}
  C.~Zhang, J.~Lewandowski, H.~Li, Y.~Ma,
  {Bouncing evolution in a model of loop quantum gravity}.
  Phys. Rev. D \textbf{99}, 124012 (2019).
  {\url{https://doi.org/10.1103/PhysRevD.99.124012}}.
  {\href{https://arxiv.org/abs/1904.07046}{{arXiv:1904.07046}}}

\bibitem{Engle:2007wy}
  J.~Engle, E.~Livine, R.~Pereira, C.~Rovelli,
  {LQG vertex with finite Immirzi parameter}.
  Nucl. Phys. B \textbf{799}, 136 (2008).
  {\url{https://doi.org/10.1016/j.nuclphysb.2008.02.018}}.
  {\href{https://arxiv.org/abs/0711.0146}{{arXiv:0711.0146}}}

\bibitem{Freidel:2007py}
  L.~Freidel, K.~Krasnov,
  {A new spin foam model for 4D gravity}.
  Class. Quant. Grav. \textbf{25}, 125018 (2008).
  {\url{https://doi.org/10.1088/0264-9381/25/12/125018}}.
  {\href{https://arxiv.org/abs/0708.1595}{{arXiv:0708.1595}}}

\bibitem{Alesci:2011ia}
  E.~Alesci, T.~Thiemann, A.~Zipfel,
  {Linking covariant and canonical LQG: New solutions to the Euclidean scalar
  constraint}.
  Phys. Rev. D \textbf{86}, 024017 (2012).
  {\url{https://doi.org/10.1103/PhysRevD.86.024017}}.
  {\href{https://arxiv.org/abs/1109.1290}{{arXiv:1109.1290}}}

\bibitem{Yang:2021den}
  J.~Yang, C.~Zhang, Y.~Ma,
  {Relating spin-foam to canonical loop quantum gravity by graphical calculus}.
  Phys. Rev. D \textbf{104}, 044025 (2021).
  {\url{https://doi.org/10.1103/PhysRevD.104.044025}}.
  {\href{https://arxiv.org/abs/2102.05881}{{arXiv:2102.05881}}}

\bibitem{Lewandowski:2021bkt}
  J.~Lewandowski, C.~Zhang,
  {Fermion coupling to loop quantum gravity: Canonical formulation}.
  Phys. Rev. D \textbf{105}, 124025 (2022).
  {\url{https://doi.org/10.1103/PhysRevD.105.124025}}.
  {\href{https://arxiv.org/abs/2112.08865}{{arXiv:2112.08865}}}

\bibitem{Zhang:2022bzp}
  C.~Zhang, H.~Liu, M.~Han,
  {Fermions on quantum geometry and resolution of doubling problem}.
  {\href{https://arxiv.org/abs/2205.12208}{{arXiv:2205.12208}}}

\bibitem{Bodendorfer:2011nx}
  N.~Bodendorfer, T.~Thiemann, A.~Thurn,
  {New variables for classical and quantum gravity in all dimensions: III.
  Quantum theory}.
  Class. Quant. Grav. \textbf{30}, 045003 (2013).
  {\url{https://doi.org/10.1088/0264-9381/30/4/045003}}.
  {\href{https://arxiv.org/abs/1105.3705}{{arXiv:1105.3705}}}

\bibitem{Han:2013noa}
  Y.~Han, Y.~Ma, X.~Zhang,
  {Connection dynamics for higher dimensional scalar-tensor theories of
  gravity}.
  Mod. Phys. Lett. A \textbf{29}, 1450134 (2014).
  {\url{https://doi.org/10.1142/S021773231450134X}}.
  {\href{https://arxiv.org/abs/1304.0209}{{arXiv:1304.0209}}}

\bibitem{Long:2019nkf}
  G.~Long, C.Y. Lin, Y.~Ma,
  {Coherent intertwiner solution of simplicity constraint in all dimensional
  loop quantum gravity}.
  Phys. Rev. D \textbf{100}, 064065 (2019).
  {\url{https://doi.org/10.1103/PhysRevD.100.064065}}.
  {\href{https://arxiv.org/abs/1906.06534}{{arXiv:1906.06534}}}

\bibitem{Long:2020wuj}
  G.~Long, Y.~Ma,
  {General geometric operators in all dimensional loop quantum gravity}.
  Phys. Rev. D \textbf{101}, 084032 (2020).
  {\url{https://doi.org/10.1103/PhysRevD.101.084032}}.
  {\href{https://arxiv.org/abs/2003.03952}{{arXiv:2003.03952}}}

\bibitem{Long:2020agv}
  G.~Long, Y.~Ma,
  {Polytopes in all dimensional loop quantum gravity}.
  Eur. Phys. J. C \textbf{82}, 41 (2022).
  {\url{https://doi.org/10.1140/epjc/s10052-022-09988-2}}.
  {\href{https://arxiv.org/abs/2009.11196}{{arXiv:2009.11196}}}

\bibitem{Zhang:2011vi}
  X.~Zhang, Y.~Ma,
  {Extension of Loop Quantum Gravity to $f(R)$ Theories}.
  Phys. Rev. Lett. \textbf{106}, 171301 (2011).
  {\url{https://doi.org/10.1103/PhysRevLett.106.171301}}.
  {\href{https://arxiv.org/abs/1101.1752}{{arXiv:1101.1752}}}

\bibitem{Zhang:2011qq}
  X.~Zhang, Y.~Ma,
  {Loop quantum $f(R)$ theories}.
  Phys. Rev. D \textbf{84}, 064040 (2011).
  {\url{https://doi.org/10.1103/PhysRevD.84.064040}}.
  {\href{https://arxiv.org/abs/1107.4921}{{arXiv:1107.4921}}}

\bibitem{Zhang:2011vg}
  X.~Zhang, Y.~Ma,
  {Nonperturbative loop quantization of scalar-tensor theories of gravity}.
  Phys. Rev. D \textbf{84}, 104045 (2011).
  {\url{https://doi.org/10.1103/PhysRevD.84.104045}}.
  {\href{https://arxiv.org/abs/1107.5157}{{arXiv:1107.5157}}}

\bibitem{Zhang:2011gn}
  X.~Zhang, Y.~Ma,
  {Loop quantum Brans-Dicke theory}.
  J. Phys. Conf. Ser. \textbf{360}, 012055 (2012).
  {\url{https://doi.org/10.1088/1742-6596/360/1/012055}}.
  {\href{https://arxiv.org/abs/1111.2215}{{arXiv:1111.2215}}}

\bibitem{Ma:2011aa}
  Y.~Ma,
  {Extension of loop quantum gravity to metric theories beyond general
  relativity}.
  J. Phys. Conf. Ser. \textbf{360}, 012006 (2012).
  {\url{https://doi.org/10.1088/1742-6596/360/1/012006}}.
  {\href{https://arxiv.org/abs/1112.2085}{{arXiv:1112.2085}}}

\bibitem{Zhou:2012ie}
  Z.~Zhou, H.~Guo, Y.~Han, Y.~Ma,
  {Action principle for the connection dynamics of scalar-tensor theories}.
  Phys. Rev. D \textbf{87}, 087502 (2013).
  {\url{https://doi.org/10.1103/PhysRevD.87.087502}}.
  {\href{https://arxiv.org/abs/1211.5939}{{arXiv:1211.5939}}}

\bibitem{Chen:2018dqz}
  Q.~Chen, Y.~Ma,
  {Hamiltonian structure and connection-dynamics of Weyl gravity}.
  Phys. Rev. D \textbf{98}, 064009 (2018).
  {\url{https://doi.org/10.1103/PhysRevD.98.064009}}.
  {\href{https://arxiv.org/abs/1803.10807}{{arXiv:1803.10807}}}

\bibitem{Zhang:2020smo}
  X.~Zhang, J.~Yang, Y.~Ma,
  {Canonical loop quantization of the lowest-order projectable Horava gravity}.
  Phys. Rev. D \textbf{102}, 124060 (2020).
  {\url{https://doi.org/10.1103/PhysRevD.102.124060}}.
  {\href{https://arxiv.org/abs/2008.04553}{{arXiv:2008.04553}}}

\bibitem{Ashtekar:2003hd}
  A.~Ashtekar, M.~Bojowald, J.~Lewandowski,
  {Mathematical structure of loop quantum cosmology}.
  Adv. Theor. Math. Phys. \textbf{7}, 233 (2003).
  {\url{https://doi.org/10.4310/ATMP.2003.v7.n2.a2}}.
  {\href{https://arxiv.org/abs/gr-qc/0304074}{{arXiv:gr-qc/0304074}}}

\bibitem{Ashtekar:2006rx}
  A.~Ashtekar, T.~Pawlowski, P.~Singh,
  {Quantum nature of the big bang}.
  Phys. Rev. Lett. \textbf{96}, 141301 (2006).
  {\url{https://doi.org/10.1103/PhysRevLett.96.141301}}.
  {\href{https://arxiv.org/abs/gr-qc/0602086}{{arXiv:gr-qc/0602086}}}

\bibitem{Ashtekar:2006wn}
  A.~Ashtekar, T.~Pawlowski, P.~Singh,
  {Quantum nature of the big bang: Improved dynamics}.
  Phys. Rev. D \textbf{74}, 084003 (2006).
  {\url{https://doi.org/10.1103/PhysRevD.74.084003}}.
  {\href{https://arxiv.org/abs/gr-qc/0607039}{{arXiv:gr-qc/0607039}}}

\bibitem{Ding:2008tq}
  Y.~Ding, Y.~Ma, J.~Yang,
  {Effective Scenario of Loop Quantum Cosmology}.
  Phys. Rev. Lett. \textbf{102}, 051301 (2009).
  {\url{https://doi.org/10.1103/PhysRevLett.102.051301}}.
  {\href{https://arxiv.org/abs/0808.0990}{{arXiv:0808.0990}}}

\bibitem{Yang:2009fp}
  J.~Yang, Y.~Ding, Y.~Ma,
  {Alternative quantization of the Hamiltonian in loop quantum cosmology}.
  Phys. Lett. B \textbf{682}, 1 (2009).
  {\url{https://doi.org/10.1016/j.physletb.2009.10.072}}.
  {\href{https://arxiv.org/abs/0904.4379}{{arXiv:0904.4379}}}

\bibitem{Assanioussi:2018hee}
  M.~Assanioussi, A.~Dapor, K.~Liegener, T.~Paw\l{}owski,
  {Emergent de Sitter Epoch of the Quantum Cosmos from Loop Quantum Cosmology}.
  Phys. Rev. Lett. \textbf{121}, 081303 (2018).
  {\url{https://doi.org/10.1103/PhysRevLett.121.081303}}.
  {\href{https://arxiv.org/abs/1801.00768}{{arXiv:1801.00768}}}

\bibitem{Li:2018opr}
  B.F. Li, P.~Singh, A.~Wang,
  {Towards cosmological dynamics from loop quantum gravity}.
  Phys. Rev. D \textbf{97}, 084029 (2018).
  {\url{https://doi.org/10.1103/PhysRevD.97.084029}}.
  {\href{https://arxiv.org/abs/1801.07313}{{arXiv:1801.07313}}}

\bibitem{Ashtekar:2005qt}
  A.~Ashtekar, M.~Bojowald,
  {Quantum geometry and the Schwarzschild singularity}.
  Class. Quant. Grav. \textbf{23}, 391 (2006).
  {\url{https://doi.org/10.1088/0264-9381/23/2/008}}.
  {\href{https://arxiv.org/abs/gr-qc/0509075}{{arXiv:gr-qc/0509075}}}

\bibitem{Gambini:2020nsf}
  R.~Gambini, J.~Olmedo, J.~Pullin,
  {Spherically symmetric loop quantum gravity: Analysis of improved dynamics}.
  Class. Quant. Grav. \textbf{37}, 205012 (2020).
  {\url{https://doi.org/10.1088/1361-6382/aba842}}.
  {\href{https://arxiv.org/abs/2006.01513}{{arXiv:2006.01513}}}

\bibitem{Han:2019vpw}
  M.~Han, H.~Liu,
  {Effective dynamics from coherent state path integral of full loop quantum
  gravity}.
  Phys.\ Rev.\ D \textbf{101}, 046003 (2020).
  {\url{https://doi.org/10.1103/PhysRevD.101.046003}}.
  {\href{https://arxiv.org/abs/1910.03763}{{arXiv:1910.03763}}}

\bibitem{Han:2019feb}
  M.~Han, H.~Liu,
  {Improved $\bar{\mu}$-scheme effective dynamics of full loop quantum
  gravity}.
  Phys. Rev. D \textbf{102}, 064061 (2020).
  {\url{https://doi.org/10.1103/PhysRevD.102.064061}}.
  {\href{https://arxiv.org/abs/1912.08668}{{arXiv:1912.08668}}}

\bibitem{Liegener:2020dcg}
  K.~Liegener, L.~Rudnicki,
  {Algorithmic approach to cosmological coherent state expectation values in
  loop quantum gravity}.
  Class. Quant. Grav. \textbf{38}, 205001 (2021).
  {\url{https://doi.org/10.1088/1361-6382/ac226f}}.
  {\href{https://arxiv.org/abs/2012.07813}{{arXiv:2012.07813}}}

\bibitem{Zhang:2020mld}
  C.~Zhang, S.~Song, M.~Han,
  {First-order quantum correction in coherent state expectation value of
  loop-quantum-gravity Hamiltonian: Overview and results}.
  {\href{https://arxiv.org/abs/2012.14242}{{arXiv:2012.14242}}}

\bibitem{Zhang:2021qul}
  C.~Zhang, S.~Song, M.~Han,
  {First-order quantum correction in coherent state expectation value of
  loop-quantum-gravity Hamiltonian}.
  Phys. Rev. D \textbf{105}, 064008 (2022).
  {\url{https://doi.org/10.1103/PhysRevD.105.064008}}.
  {\href{https://arxiv.org/abs/2102.03591}{{arXiv:2102.03591}}}

\bibitem{Oppenheimer:1939ue}
  J.R. Oppenheimer, H.~Snyder,
  {On continued gravitational contraction}.
  Phys. Rev. \textbf{56}, 455 (1939).
  {\url{https://doi.org/10.1103/PhysRev.56.455}}

\bibitem{Bojowald:2005qw}
  M.~Bojowald, R.~Goswami, R.~Maartens, P.~Singh,
  {Black Hole Mass Threshold from Nonsingular Quantum Gravitational Collapse}.
  Phys. Rev. Lett. \textbf{95}, 091302 (2005).
  {\url{https://doi.org/10.1103/PhysRevLett.95.091302}}.
  {\href{https://arxiv.org/abs/gr-qc/0503041}{{arXiv:gr-qc/0503041}}}

\bibitem{Bojowald:2009ih}
  M.~Bojowald, J.D. Reyes, R.~Tibrewala,
  {Non-marginal LTB-like models with inverse triad corrections from loop
  quantum gravity}.
  Phys. Rev. D \textbf{80}, 084002 (2009).
  {\url{https://doi.org/10.1103/PhysRevD.80.084002}}.
  {\href{https://arxiv.org/abs/0906.4767}{{arXiv:0906.4767}}}

\bibitem{Marto:2013soa}
  J.~Marto, Y.~Tavakoli, P.~Vargas~Moniz,
  {Improved dynamics and gravitational collapse of tachyon field coupled with a
  barotropic fluid}.
  Int. J. Mod. Phys. D \textbf{24}, 1550025 (2015).
  {\url{https://doi.org/10.1142/S021827181550025X}}.
  {\href{https://arxiv.org/abs/1308.4953}{{arXiv:1308.4953}}}

\bibitem{Kelly:2020lec}
  J.G. Kelly, R.~Santacruz, E.~Wilson-Ewing,
  {Black hole collapse and bounce in effective loop quantum gravity}.
  Class. Quant. Grav. \textbf{38}, 04LT01 (2021).
  {\url{https://doi.org/10.1088/1361-6382/abd3e2}}.
  {\href{https://arxiv.org/abs/2006.09325}{{arXiv:2006.09325}}}

\bibitem{BenAchour:2020bdt}
  J.~Ben~Achour, S.~Brahma, J.P. Uzan,
  {Bouncing compact objects. Part I. Quantum extension of the
  Oppenheimer-Snyder collapse}.
  J. Cosmol. Astropart. Phys. \textbf{03}, 041 (2020).
  {\url{https://doi.org/10.1088/1475-7516/2020/03/041}}.
  {\href{https://arxiv.org/abs/2001.06148}{{arXiv:2001.06148}}}

\bibitem{BenAchour:2020gon}
  J.~Ben~Achour, S.~Brahma, S.~Mukohyama, J.P. Uzan,
  {Towards consistent black-to-white hole bounces from matter collapse}.
  J. Cosmol. Astropart. Phys. \textbf{09}, 020 (2020).
  {\url{https://doi.org/10.1088/1475-7516/2020/09/020}}.
  {\href{https://arxiv.org/abs/2004.12977}{{arXiv:2004.12977}}}

\bibitem{Munch:2020czs}
  J.~M\"unch,
  {Effective quantum dust collapse via surface matching}.
  Class. Quant. Grav. \textbf{38}, 175015 (2021).
  {\url{https://doi.org/10.1088/1361-6382/ac103e}}.
  {\href{https://arxiv.org/abs/2010.13480}{{arXiv:2010.13480}}}

\bibitem{Munch:2021oqn}
  J.~M\"unch,
  {Causal structure of a recent loop quantum gravity black hole collapse
  model}.
  Phys. Rev. D \textbf{104}, 046019 (2021).
  {\url{https://doi.org/10.1103/PhysRevD.104.046019}}.
  {\href{https://arxiv.org/abs/2103.17112}{{arXiv:2103.17112}}}

\bibitem{Husain:2021ojz}
  V.~Husain, J.G. Kelly, R.~Santacruz, E.~Wilson-Ewing,
  {Quantum Gravity of Dust Collapse: Shock Waves from Black Holes}.
  Phys. Rev. Lett. \textbf{128}, 121301 (2022).
  {\url{https://doi.org/10.1103/PhysRevLett.128.121301}}.
  {\href{https://arxiv.org/abs/2109.08667}{{arXiv:2109.08667}}}

\bibitem{Giesel:2021dug}
  K.~Giesel, B.F. Li, P.~Singh,
  {Nonsingular quantum gravitational dynamics of an Lema\^\i{}tre-Tolman-Bondi
  dust shell model: The role of quantization prescriptions}.
  Phys. Rev. D \textbf{104}, 106017 (2021).
  {\url{https://doi.org/10.1103/PhysRevD.104.106017}}.
  {\href{https://arxiv.org/abs/2107.05797}{{arXiv:2107.05797}}}

\bibitem{Lewandowski:2022zce}
  J.~Lewandowski, Y.~Ma, J.~Yang, C.~Zhang,
  {Quantum Oppenheimer-Snyder and Swiss Cheese Models}.
  Phys. Rev. Lett. \textbf{130}, 101501 (2023).
  {\url{https://doi.org/10.1103/PhysRevLett.130.101501}}.
  {\href{https://arxiv.org/abs/2210.02253}{{arXiv:2210.02253}}}

\bibitem{LIGOScientific:2017bnn}
  B.P. Abbott {et~al.}, (LIGO Scientific Collaboration and VIRGO
  Collaboration),
  {GW170104: Observation of a 50-Solar-Mass Binary Black Hole Coalescence at
  Redshift 0.2}.
  Phys. Rev. Lett. \textbf{118}, 221101 (2017).
  {\url{https://doi.org/10.1103/PhysRevLett.118.221101}}.
  {\href{https://arxiv.org/abs/1706.01812}{{arXiv:1706.01812}}}

\bibitem{EventHorizonTelescope:2019dse}
  K.~Akiyama {et~al.}, (Event Horizon Telescope Collaboration),
  {First M87 Event Horizon Telescope Results. I. The Shadow of the Supermassive
  Black Hole}.
  Astrophys. J. Lett. \textbf{875}, L1 (2019).
  {\url{https://doi.org/10.3847/2041-8213/ab0ec7}}.
  {\href{https://arxiv.org/abs/1906.11238}{{arXiv:1906.11238}}}

\bibitem{EventHorizonTelescope:2022wkp}
  K.~Akiyama {et~al.}, (Event Horizon Telescope Collaboration),
  {First Sagittarius A* Event Horizon Telescope Results. I. The Shadow of the
  Supermassive Black Hole in the Center of the Milky Way}.
  Astrophys. J. Lett. \textbf{930}, L12 (2022).
  {\url{https://doi.org/10.3847/2041-8213/ac6674}}

\bibitem{Bardeen:1973}
  J.M. Bardeen,
  Timelike and null geodesics in the kerr metric,
  in \emph{{Black Holes (Les Astres Occlus)}},
  edited by C.~DeWitt, B.S. DeWitt
  (Gordon and Breach Science Publishers, Inc., New York, 1973)

\bibitem{Gralla:2019xty}
  S.E. Gralla, D.E. Holz, R.M. Wald,
  {Black hole shadows, photon rings, and lensing rings}.
  Phys. Rev. D \textbf{100}, 024018 (2019).
  {\url{https://doi.org/10.1103/PhysRevD.100.024018}}.
  {\href{https://arxiv.org/abs/1906.00873}{{arXiv:1906.00873}}}

\bibitem{Regge:1957td}
  T.~Regge, J.A. Wheeler,
  {Stability of a Schwarzschild singularity}.
  Phys. Rev. \textbf{108}, 1063 (1957).
  {\url{https://doi.org/10.1103/PhysRev.108.1063}}

\bibitem{Chandrasekhar:1975zza}
  S.~Chandrasekhar, S.L. Detweiler,
  {The quasi-normal modes of the Schwarzschild black hole}.
  Proc. Roy. Soc. A \textbf{344}, 441 (1975).
  {\url{https://doi.org/10.1098/rspa.1975.0112}}

\bibitem{Gundlach:1993tp}
  C.~Gundlach, R.H. Price, J.~Pullin,
  {Late-time behavior of stellar collapse and explosions. I. Linearized
  perturbations}.
  Phys. Rev. D \textbf{49}, 883 (1994).
  {\url{https://doi.org/10.1103/PhysRevD.49.883}}.
  {\href{https://arxiv.org/abs/gr-qc/9307009}{{arXiv:gr-qc/9307009}}}

\bibitem{Zhang:2023okw}
  C.~Zhang, Y.~Ma, J.~Yang,
  {Black hole image encoding quantum gravity information}.
  {\href{https://arxiv.org/abs/2302.02800}{{arXiv:2302.02800}}}

\bibitem{Darmois:1927}
  G.~Darmois,
  {Les \'equations de la gravitation einsteinienne},
  in \emph{{M\'emorial des Sciences Math\'ematiques}}
  (Gauthier-Villars, Paris, 1927)

\bibitem{Israel:1966rt}
  W.~Israel,
  {Singular hypersurfaces and thin shells in general relativity}.
  Nuovo Cim. B \textbf{44S10}, 1 (1966).
  {\url{https://doi.org/10.1007/BF02710419}}

\bibitem{Poisson:2004bk}
  E.~Poisson, \emph{{A Relativist's Toolkit: The Mathematics of Black-Hole
  Mechanics}}
  (Cambridge University Press, Cambridge, England, 2004)

\bibitem{Piechocki:2020bfo}
  W.~Piechocki, T.~Schmitz,
  {Quantum Oppenheimer-Snyder model}.
  Phys. Rev. D \textbf{102}, 046004 (2020).
  {\url{https://doi.org/10.1103/PhysRevD.102.046004}}.
  {\href{https://arxiv.org/abs/2004.02939}{{arXiv:2004.02939}}}

\bibitem{Vandersloot:2006ws}
  K.~Vandersloot,
  {Loop quantum cosmology and the $k=-1$ Robertson-Walker model}.
  Phys. Rev. D \textbf{75}, 023523 (2007).
  {\url{https://doi.org/10.1103/PhysRevD.75.023523}}.
  {\href{https://arxiv.org/abs/gr-qc/0612070}{{arXiv:gr-qc/0612070}}}

\bibitem{Vandersloot:2006gga}
  K.~Vandersloot,
 \emph{{Loop Quantum Cosmology}}
 (Ph.D. thesis, The Pennsylvania State University, 2006)

\bibitem{Domagala:2004jt}
  M.~Domagala, J.~Lewandowski,
  {Black hole entropy from quantum geometry}.
  Class. Quant. Grav. \textbf{21}, 5233 (2004).
  {\url{https://doi.org/10.1088/0264-9381/21/22/014}}.
  {\href{https://arxiv.org/abs/gr-qc/0407051}{{arXiv:gr-qc/0407051}}}

\bibitem{Meissner:2004ju}
  K.A. Meissner,
  {Black hole entropy in loop quantum gravity}.
  Class. Quant. Grav. \textbf{21}, 5245 (2004).
  {\url{https://doi.org/10.1088/0264-9381/21/22/015}}.
  {\href{https://arxiv.org/abs/gr-qc/0407052}{{arXiv:gr-qc/0407052}}}

\bibitem{Kelly:2020uwj}
  J.G. Kelly, R.~Santacruz, E.~Wilson-Ewing,
  {Effective loop quantum gravity framework for vacuum spherically symmetric
  spacetimes}.
  Phys. Rev. D \textbf{102}, 106024 (2020).
  {\url{https://doi.org/10.1103/PhysRevD.102.106024}}.
  {\href{https://arxiv.org/abs/2006.09302}{{arXiv:2006.09302}}}

\bibitem{Peng:2020wun}
  J.~Peng, M.~Guo, X.H. Feng,
  {Influence of quantum correction on black hole shadows, photon rings, and
  lensing rings}.
  Chin. Phys. C \textbf{45}, 085103 (2021).
  {\url{https://doi.org/10.1088/1674-1137/ac06bb}}.
  {\href{https://arxiv.org/abs/2008.00657}{{arXiv:2008.00657}}}

\bibitem{Lu:2019zxb}
  H.~Lu, H.D. Lyu,
  {Schwarzschild black holes have the largest size}.
  Phys. Rev. D \textbf{101}, 044059 (2020).
  {\url{https://doi.org/10.1103/PhysRevD.101.044059}}.
  {\href{https://arxiv.org/abs/1911.02019}{{arXiv:1911.02019}}}

\bibitem{Feng:2019zzn}
  X.H. Feng, H.~Lu,
  {On the size of rotating black holes}.
  Eur. Phys. J. C \textbf{80}, 551 (2020).
  {\url{https://doi.org/10.1140/epjc/s10052-020-8119-z}}.
  {\href{https://arxiv.org/abs/1911.12368}{{arXiv:1911.12368}}}

\bibitem{Kokkotas:1999bd}
  K.D. Kokkotas, B.G. Schmidt,
  {Quasinormal modes of stars and black holes}.
  Living Rev. Rel. \textbf{2}, 2 (1999).
  {\url{https://doi.org/10.12942/lrr-1999-2}}.
  {\href{https://arxiv.org/abs/gr-qc/9909058}{{arXiv:gr-qc/9909058}}}

\bibitem{Konoplya:2011qq}
  R.A. Konoplya, A.~Zhidenko,
  {Quasinormal modes of black holes: From astrophysics to string theory}.
  Rev. Mod. Phys. \textbf{83}, 793 (2011).
  {\url{https://doi.org/10.1103/RevModPhys.83.793}}.
  {\href{https://arxiv.org/abs/1102.4014}{{arXiv:1102.4014}}}

\bibitem{Zerilli:1970se}
  F.J. Zerilli,
  {Effective Potential for Even-Parity Regge-Wheeler Gravitational Perturbation
  Equations}.
  Phys. Rev. Lett. \textbf{24}, 737 (1970).
  {\url{https://doi.org/10.1103/PhysRevLett.24.737}}

\bibitem{Schutz:1985km}
  B.F. Schutz, C.M. Will,
  {Black hole normal modes: A seminalytic approach}.
  Astrophys. J. Lett. \textbf{291}, L33 (1985).
  {\url{https://doi.org/10.1086/184453}}

\bibitem{Iyer:1986np}
  S.~Iyer, C.M. Will,
  {Black-hole normal modes: A WKB approach. I. Foundations and application of a
  higher-order WKB analysis of potential-barrier scattering}.
  Phys. Rev. D \textbf{35}, 3621 (1987).
  {\url{https://doi.org/10.1103/PhysRevD.35.3621}}

\bibitem{Wang:2000gsa}
  B.~Wang, C.Y. Lin, E.~Abdalla,
  {Quasinormal modes of Reissner-Nordstrom anti-de Sitter black holes}.
  Phys. Lett. B \textbf{481}, 79 (2000).
  {\url{https://doi.org/10.1016/S0370-2693(00)00409-3}}.
  {\href{https://arxiv.org/abs/hep-th/0003295}{{arXiv:hep-th/0003295}}}

\bibitem{Li:2015mqa}
  R.~Li, Y.~Tian, H.~Zhang, J.~Zhao,
  {Time domain analysis of superradiant instability for the charged stringy
  black hole\textendash{}mirror system}.
  Phys. Lett. B \textbf{750}, 520 (2015).
  {\url{https://doi.org/10.1016/j.physletb.2015.09.073}}.
  {\href{https://arxiv.org/abs/1506.04267}{{arXiv:1506.04267}}}

\bibitem{Zou:2017juz}
  D.C. Zou, Y.~Liu, R.H. Yue,
  {Behavior of quasinormal modes and Van der Waals-like phase transition of
  charged AdS black holes in massive gravity}.
  Eur. Phys. J. C \textbf{77}, 365 (2017).
  {\url{https://doi.org/10.1140/epjc/s10052-017-4937-z}}.
  {\href{https://arxiv.org/abs/1702.08118}{{arXiv:1702.08118}}}

\bibitem{Zhang:2020sjh}
  C.Y. Zhang, S.J. Zhang, P.C. Li, M.~Guo,
  {Superradiance and stability of the regularized 4D charged
  Einstein-Gauss-Bonnet black hole}.
  J. High Energy Phys. \textbf{08}, 105 (2020).
  {\url{https://doi.org/10.1007/JHEP08(2020)105}}.
  {\href{https://arxiv.org/abs/2004.03141}{{arXiv:2004.03141}}}

\bibitem{Qian:2020wbv}
  W.L. Qian, K.~Lin, J.P. Wu, B.~Wang, R.H. Yue,
  {On quasinormal frequencies of black hole perturbations with an external
  source}.
  Eur. Phys. J. C \textbf{80}, 959 (2020).
  {\url{https://doi.org/10.1140/epjc/s10052-020-08539-x}}.
  {\href{https://arxiv.org/abs/2006.07122}{{arXiv:2006.07122}}}

\bibitem{Liu:2020evp}
  P.~Liu, C.~Niu, C.Y. Zhang,
  {Instability of regularized 4D charged Einstein-Gauss-Bonnet de-Sitter black
  holes}.
  Chin. Phys. C \textbf{45}, 025104 (2021).
  {\url{https://doi.org/10.1088/1674-1137/abcd2d}}.
  {\href{https://arxiv.org/abs/2004.10620}{{arXiv:2004.10620}}}

\bibitem{Liu:2021fzr}
  P.~Liu, C.~Niu, C.Y. Zhang,
  {Linear instability of charged massless scalar perturbation in regularized 4D
  charged Einstein-Gauss-Bonnet anti de-Sitter black holes}.
  Chin. Phys. C \textbf{45}, 025111 (2021).
  {\url{https://doi.org/10.1088/1674-1137/abd01d}}

\bibitem{Wang:2021upj}
  M.~Wang, Z.~Chen, X.~Tong, Q.~Pan, J.~Jing,
  {Bifurcation of the Maxwell quasinormal spectrum on asymptotically
  anti\textendash{}de Sitter black holes}.
  Phys. Rev. D \textbf{103}, 064079 (2021).
  {\url{https://doi.org/10.1103/PhysRevD.103.064079}}.
  {\href{https://arxiv.org/abs/2104.04970}{{arXiv:2104.04970}}}

\bibitem{Wang:2021uix}
  M.~Wang, Z.~Chen, Q.~Pan, J.~Jing,
  {Maxwell quasinormal modes on a global monopole Schwarzschild-anti-de Sitter
  black hole with Robin boundary conditions}.
  Eur. Phys. J. C \textbf{81}, 469 (2021).
  {\url{https://doi.org/10.1140/epjc/s10052-021-09149-x}}.
  {\href{https://arxiv.org/abs/2105.10951}{{arXiv:2105.10951}}}

\bibitem{Dreyer:2002vy}
  O.~Dreyer,
  {Quasinormal Modes, the Area Spectrum, and Black Hole Entropy}.
  Phys. Rev. Lett. \textbf{90}, 081301 (2003).
  {\url{https://doi.org/10.1103/PhysRevLett.90.081301}}.
  {\href{https://arxiv.org/abs/gr-qc/0211076}{{arXiv:gr-qc/0211076}}}

\bibitem{Santos:2015gja}
  V.~Santos, R.V. Maluf, C.A.S. Almeida,
  {Quasinormal frequencies of self-dual black holes}.
  Phys. Rev. D \textbf{93}, 084047 (2016).
  {\url{https://doi.org/10.1103/PhysRevD.93.084047}}.
  {\href{https://arxiv.org/abs/1509.04306}{{arXiv:1509.04306}}}

\bibitem{Cruz:2015bcj}
  M.B. Cruz, C.A.S. Silva, F.A. Brito,
  {Gravitational axial perturbations and quasinormal modes of loop quantum
  black holes}.
  Eur. Phys. J. C \textbf{79}, 157 (2019).
  {\url{https://doi.org/10.1140/epjc/s10052-019-6565-2}}.
  {\href{https://arxiv.org/abs/1511.08263}{{arXiv:1511.08263}}}

\bibitem{Anacleto:2020zhp}
  M.A. Anacleto, F.A. Brito, J.A.V. Campos, E.~Passos,
  {Absorption and scattering by a self-dual black hole}.
  Gen. Rel. Grav. \textbf{52}, 100 (2020).
  {\url{https://doi.org/10.1007/s10714-020-02756-1}}.
  {\href{https://arxiv.org/abs/2002.12090}{{arXiv:2002.12090}}}

\bibitem{Liu:2020ola}
  C.~Liu, T.~Zhu, Q.~Wu, K.~Jusufi, M.~Jamil, M.~Azreg-A\"\i{}nou, A.~Wang,
  {Shadow and quasinormal modes of a rotating loop quantum black hole}.
  Phys. Rev. D \textbf{101}, 084001 (2020).
  {\url{https://doi.org/10.1103/PhysRevD.101.084001}}.
  {\href{https://arxiv.org/abs/2003.00477}{{arXiv:2003.00477}}}

\bibitem{Bouhmadi-Lopez:2020oia}
  M.~Bouhmadi-L\'opez, S.~Brahma, C.Y. Chen, P.~Chen, D.h. Yeom,
  {A consistent model of non-singular Schwarzschild black hole in loop quantum
  gravity and its quasinormal modes}.
  J. Cosmol. Astropart. Phys. \textbf{07}, 066 (2020).
  {\url{https://doi.org/10.1088/1475-7516/2020/07/066}}.
  {\href{https://arxiv.org/abs/2004.13061}{{arXiv:2004.13061}}}

\bibitem{Cruz:2020emz}
  M.B. Cruz, F.A. Brito, C.A.S. Silva,
  {Polar gravitational perturbations and quasinormal modes of a loop quantum
  gravity black hole}.
  Phys. Rev. D \textbf{102}, 044063 (2020).
  {\url{https://doi.org/10.1103/PhysRevD.102.044063}}.
  {\href{https://arxiv.org/abs/2005.02208}{{arXiv:2005.02208}}}

\bibitem{Daghigh:2020fmw}
  R.G. Daghigh, M.D. Green, G.~Kunstatter,
  {Scalar perturbations and stability of a loop quantum corrected Kruskal black
  hole}.
  Phys. Rev. D \textbf{103}, 084031 (2021).
  {\url{https://doi.org/10.1103/PhysRevD.103.084031}}.
  {\href{https://arxiv.org/abs/2012.13359}{{arXiv:2012.13359}}}

\bibitem{Santos:2021wsw}
  J.S. Santos, M.B. Cruz, F.A. Brito,
  {Quasinormal modes of a massive scalar field nonminimally coupled to gravity
  in the spacetime of self-dual black hole}.
  Eur. Phys. J. C \textbf{81}, 1082 (2021).
  {\url{https://doi.org/10.1140/epjc/s10052-021-09884-1}}.
  {\href{https://arxiv.org/abs/2103.11212}{{arXiv:2103.11212}}}

\bibitem{Liu:2021djf}
  Y.C. Liu, J.X. Feng, F.W. Shu, A.~Wang,
  {Extended geometry of Gambini-Olmedo-Pullin polymer black hole and its
  quasinormal spectrum}.
  Phys. Rev. D \textbf{104}, 106001 (2021).
  {\url{https://doi.org/10.1103/PhysRevD.104.106001}}.
  {\href{https://arxiv.org/abs/2109.02861}{{arXiv:2109.02861}}}

\bibitem{del-Corral:2022kbk}
  D.~del Corral, J.~Olmedo,
  {Breaking of isospectrality of quasinormal modes in nonrotating loop quantum
  gravity black holes}.
  Phys. Rev. D \textbf{105}, 064053 (2022).
  {\url{https://doi.org/10.1103/PhysRevD.105.064053}}.
  {\href{https://arxiv.org/abs/2201.09584}{{arXiv:2201.09584}}}

\bibitem{Momennia:2022tug}
  M.~Momennia,
  {Quasinormal modes of self-dual black holes in loop quantum gravity}.
  Phys. Rev. D \textbf{106}, 024052 (2022).
  {\url{https://doi.org/10.1103/PhysRevD.106.024052}}.
  {\href{https://arxiv.org/abs/2204.03259}{{arXiv:2204.03259}}}

\bibitem{Konoplya:2003ii}
  R.A. Konoplya,
  {Quasinormal behavior of the $D$-dimensional Schwarzschild black hole and
  higher order WKB approach}.
  Phys. Rev. D \textbf{68}, 024018 (2003).
  {\url{https://doi.org/10.1103/PhysRevD.68.024018}}.
  {\href{https://arxiv.org/abs/gr-qc/0303052}{{arXiv:gr-qc/0303052}}}

\bibitem{Konoplya:2019hlu}
  R.A. Konoplya, A.~Zhidenko, A.F. Zinhailo,
  {Higher order WKB formula for quasinormal modes and grey-body factors:
  recipes for quick and accurate calculations}.
  Class. Quant. Grav. \textbf{36}, 155002 (2019).
  {\url{https://doi.org/10.1088/1361-6382/ab2e25}}.
  {\href{https://arxiv.org/abs/1904.10333}{{arXiv:1904.10333}}}

\bibitem{Marple:1987}
  S.L. Marple, \emph{Digital Spectral Analysis with Applications}
  (Prentice-Hall, New Jersey, 1987)

\bibitem{Berti:2007dg}
  E.~Berti, V.~Cardoso, J.A. Gonzalez, U.~Sperhake,
  {Mining information from binary black hole mergers: A comparison of
  estimation methods for complex exponentials in noise}.
  Phys. Rev. D \textbf{75}, 124017 (2007).
  {\url{https://doi.org/10.1103/PhysRevD.75.124017}}.
  {\href{https://arxiv.org/abs/gr-qc/0701086}{{arXiv:gr-qc/0701086}}}

\end{thebibliography}
\end{document}